\documentclass[12pt]{article}

\pdfoutput=1

\usepackage{amsmath,amssymb,amsfonts,amscd,mathrsfs}
\usepackage[dvipsnames]{xcolor}
\definecolor{darkblue}{rgb}{0.1,0.1,.7}

\usepackage[square, comma, compress,numbers]{natbib}
\usepackage{graphicx}
\usepackage{geometry}
\usepackage[margin=10pt,font=small,labelfont=bf]{caption}
\usepackage{ifthen}

\usepackage{setspace}
\newcommand{\la}{\langle}
\newcommand{\ra}{\rangle}
\providecommand{\abs}[1]{\lvert#1\rvert}

\usepackage{tikz-feynman}
 \tikzfeynmanset{compat=1.0.0}
\usetikzlibrary[snakes]
 
 \usepackage{cancel}

\usepackage{subcaption}
\usepackage{booktabs,multirow}
\usepackage{hhline}
\usepackage{tablefootnote}
\usetikzlibrary{arrows.meta,bending,matrix,positioning}

\usepackage{tablefootnote,colortbl}
\definecolor{Gray1}{gray}{0.97}
\definecolor{Gray2}{gray}{0.9}
\definecolor{LightCyan}{rgb}{0.88,1,1}
\definecolor{blu}{rgb}{0,0,1}

\usepackage{ragged2e}
\usepackage{array}
\newcolumntype{L}[1]{>{\raggedright\let\newline\\\arraybackslash\hspace{0pt}}m{#1}}
\newcolumntype{C}[1]{>{\centering\let\newline\\\arraybackslash\hspace{0pt}}m{#1}}
\newcolumntype{R}[1]{>{\raggedleft\let\newline\\\arraybackslash\hspace{0pt}}m{#1}}

\usepackage{dsfont} 
 
\usepackage{booktabs}
\usepackage[colorlinks, linkcolor=darkblue, citecolor=darkblue, urlcolor=darkblue, linktocpage]{hyperref} 

\usepackage{titlesec}
\titleformat*{\section}{\large\bfseries}
\titleformat*{\subsection}{\normalsize\bfseries}
\titleformat*{\subsubsection}{\normalsize\it}
\titleformat*{\paragraph}{\normalsize\bfseries}
\titleformat*{\subparagraph}{\normalsize\bfseries}

\def\red{\color [rgb]{0.9,0.1,0.1}}
\def\blue{\color [rgb]{0.1,0.1,0.9}}

\newcommand{\reef}[1]{(\ref{#1})}

\def\eps{\epsilon}

\newcommand{\beq}{\begin{equation}} 
\newcommand{\eeq}{\end{equation}}
 
\def\nn{\nonumber}

\def\O {{\cal O}}

\def\geq{\geqslant}

\newcommand{\diffop}[2]{\ifthenelse{\equal{#2}{1}}{\frac{\mrm{d}}{\mrm{d} #1}}{\frac{\mrm{d}^#2}{\mrm{d} #1^#2}}}

\newcommand{\ket}[1]{|#1\rangle}
\newcommand{\bra}[1]{\langle #1|}
\newcommand{\expec}[1]{\langle #1 \rangle}

\newcommand{\mrm}[1]{{\mathrm #1}}

\usepackage[normalem]{ulem}

\catcode`,\active

\catcode`\,12

\newcommand{\colorblock}[1]{{\color{blue}#1}}
\newcommand{\colorblockr}[1]{{\color{red}#1}}
\newcommand{\colorblockp}[1]{{\color{Plum}#1}}

\newcommand{\be}{\begin{equation}}
\newcommand{\ee}{\end{equation}}
\newcommand{\bea}{\begin{eqnarray}}
\newcommand{\eea}{\end{eqnarray}}
\newcommand{\eq}[1]{Eq.~(\ref{#1})}
 
  \def\th{\theta}

\usepackage{accents}
\newlength{\dhatheight}

\newcommand{\br}[1]{\text{Br}}

\numberwithin{equation}{section}
\usepackage{tocloft}
\newcommand{\dMaster}{
 \definecolor{blue}{RGB}{0.0,0.0,255}
 \definecolor{red}{RGB}{255,0.0,0}
  \begin{minipage}[h]{0.12\linewidth}\begin{tikzpicture}
  [
roundnode/.style={circle, draw=black!60, fill=black!6, very thick, 
  inner sep=1pt,
  text width=3mm},
  squarednode/.style={rectangle, draw=gray!50, fill=gray!10, very thick, minimum size=5mm},
  squarednode2/.style={rectangle, draw=white!100, fill=white!100,   inner sep=2pt,
  text width=7.5mm}
]
\begin{feynman}[small]
\vertex (v1) at (-.2,1);
\vertex (v2) at (5,1);
\vertex (v3) at (9.7,1);
\vertex (x11) at (0,.15);
\vertex  (x12) at (0+.5,0);
\vertex (x21) at (2+.5,.15);
\vertex (x211) at (3.8+.5,.15);
\vertex (x22) at (2+.3,0);
\vertex (x221) at (3.8+.4,0);
\vertex  (x31) at (6.5+1,.15);
\vertex (x32) at (6.5+.9,0);
\vertex  (x41) at (8+1.,.15);
\vertex  (x42) at (8+1.,0);
\vertex (t2) at (3.3+.3,.58){$^{n=4\, , \  h=2}$};
\vertex  (t3) at (6.3+.93,.58){$^{n=5\, , \  h=1}$};
\vertex  (t5) at (3.5,-3.2){$^{ n=4\, , \  h=0 }$};
\vertex (t6) at (6,.2){$\scriptsize{^{\text{2-loop}}}$};
\vertex (t7) at (1.65+1.65,+1.2){$\scriptsize{^{\text{1-loop}}}$};
   \diagram*{
   (x32)-- [ very thick, red] (x221) ,
   };
\vertex (Xx32) at (5.8,0);
       \draw[->, to path={-| (\tikztotarget)}, very  thick, rounded corners, red]  (x32) --(Xx32);
   \node[squarednode] (X3) at (6.2+1,0){ $\psi^2 \phi^3$ };
   \vertex (A) at (0+.5,-.45);
   \vertex (B) at (.7,-1.5);
   \vertex (C) at (2,-2.2);
   \vertex (V1) at (-.1+.5+1.65,1);
   \vertex (V2) at (2.5+1.65,1);
   \vertex (V3) at (2.8+1.65,1);
   \vertex (V4) at (7.8,1);
   \draw[->, to path={-| (\tikztotarget)}, thick, rounded corners, blue] (V1) -- (V4);
  \filldraw[rectangle, draw=gray!50, fill=gray!10, very thick, minimum size=5mm] (2.1,.3) rectangle (4.9,-1.1);    
   \node (X2) at (3+.5,-.020){$\phi^2 F^2\ \, \ \ \ \ \, \psi^4 $    }; 
  \filldraw[fill=Apricot!20,snake=snake, segment amplitude=1] (2.85,-.35) rectangle (4.15,-.94);  
      \node (xX2) at (3.5,-.65){ $\psi^2 \phi F$   };
        \filldraw[rectangle, draw=gray!50, fill=gray!10, very thick, minimum size=5mm] (2.1,-1.5) rectangle (4.9,-2.9);    
   \node (Yy2) at (3+.5,-1.85){$\bar\psi \psi  \partial \phi^2 \ \, \,  \, \bar\psi^2 \psi^2$    }; 
      \node (Yy2) at (3.5,-2.5){$\partial^2 \phi^4$    }; 
   \vertex (D) at (4.55,-.1);
   \vertex (Dd) at (5.1,-.1);
   \vertex (E) at (5.2,-.4);
   \vertex (Ee) at (5.1,-.75);
   \vertex (F) at (4.75,-1.75);
   \draw[<-, to path={-| (\tikztotarget)},  line width = .3mm, rounded corners, blue] (D) --(Dd) -- (E) -- (Ee) -- (F);
   \vertex (D0) at (4.1,-.1);
   \vertex (E0) at (3.7,-.1);
   \vertex (F0) at (3.645,-.3);
   \draw[->, to path={-| (\tikztotarget)},  line width = .3mm, rounded corners, blue] (D0)  -- (E0)-- (F0);
   \vertex (D1) at (2.758,-.7);
      \vertex (E1) at (2.2,-.7);
      \vertex (E11) at (2.,-.9);
      \vertex (Ee1) at (2.15,-1.5);
      \vertex (F1) at (2.25,-1.75);
   \draw[<-, to path={-| (\tikztotarget)}, line width = .5mm, rounded corners, red] (D1) -- (E1) -- (E11)  -- (Ee1) --  (F1);
   \vertex (D2) at (2.58,-2.15);
   \vertex (E2) at (2.73,-2.5);
      \vertex (F2) at (3.1,-2.55);

  \end{feynman}
\end{tikzpicture}
  \end{minipage} 
  }

\newcommand{\twoloopgenUn}{
  \begin{minipage}[h]{0.12\linewidth}\begin{tikzpicture}
  [
roundnode/.style={circle, draw=black!60, fill=black!6, very thick, 
  inner sep=2.1pt,
  text width=3mm},
roundnode3/.style={circle, draw=black!60, fill=black!6, very thick, 
  inner sep=2.5pt,
  text width=3mm},
]
\begin{feynman}[small]
\vertex (X) at (-.3,0);
\vertex (Xaux) at (-.3,-.1);
\vertex (x1l) at (-1.,.4);
\vertex (xnl) at (-1.,-.4);
\vertex (x1r) at (+.3,.3);
\vertex (x2r) at (+.3,-.3);
\vertex (x3r) at (+.3,0);
\vertex (c1) at (.4,.35);
\vertex (c2) at (.4,-.4);
\node [crossed dot] (Y) at (0+1.1+.1,0);
\vertex (Yaux) at (1.08+.1,-.09);
\vertex (y1l) at (-.5+1.,.3);
\vertex (y2l) at (-.5+1.,-.3);
\vertex (y3l) at (-.5+1.,0);
\vertex (X1l) at (1.8+.1,.4);
\vertex (Xnl) at (1.8+.1,-.4);
   \diagram*{
   (x1l) -- [thick, quarter left, looseness=.8] (X) --[ thick, quarter left, looseness=.8] (x1r) ,
      (x3r) -- [thick] (X) ,
         (xnl) -- [thick, quarter right, looseness=.8] (X) --[ thick, quarter right, looseness=.8] (x2r) ,
   (y1l) -- [thick, quarter left, looseness=.8] (Y), 
     (y3l) -- [thick] (Y), 
        (y2l) -- [thick, quarter right, looseness=.8] (Y),
   (Y)   -- [thick, quarter left, looseness=.8] (X1l)  ,
    (Y)   -- [thick, quarter right, looseness=.8] (Xnl)  ,
   (c1) -- [very thick, scalar, red] (c2),
};
\node[roundnode] (Y1) at (0+1.1+.1,0){\tiny{$\vspace{-.05cm}\hspace{-.0cm}\O_i$}};
\node[roundnode3] (X1) at (-.34,0){\tiny{$\vspace{-.05cm}\hspace{-.0cm}M$}};
  \end{feynman}
\end{tikzpicture}
  \end{minipage} 
  }

\newcommand{\twoloopgenDos}{
  \begin{minipage}[h]{0.12\linewidth}\begin{tikzpicture}
  [
  roundnode/.style={circle, draw=black!60, fill=black!6, very thick, 
  inner sep=2.1pt,
  text width=3mm},
roundnode3/.style={circle, draw=black!60, fill=black!30, very thick, 
  inner sep=2.5pt,
  text width=3mm},
]
\begin{feynman}[small]
\vertex (X) at (-.3,0);
\vertex (Xaux) at (-.3,-.1);
\vertex (x1l) at (-1.,.4);
\vertex (xnl) at (-1.,-.4);
\vertex (x1r) at (+.3,.3);
\vertex (x2r) at (+.3,-.3);
\vertex (c1) at (.4,.35);
\vertex (c2) at (.4,-.4);
\node [crossed dot] (Y) at (0+1.1+.1,0);
\vertex (Yaux) at (1.08+.1,-.09);
\vertex (y1l) at (-.5+1.,.3);
\vertex (y2l) at (-.5+1.,-.3);
\vertex (X1l) at (1.8+.1,.4);
\vertex (Xnl) at (1.8+.1,-.4);
  \vertex (t7) at (-.3,-.55){$\scriptsize{^{\text{1-loop}}}$};
   \diagram*{
   (x1l) -- [thick, quarter left, looseness=.8] (X) --[ thick, quarter left, looseness=.8] (x1r) ,
         (xnl) -- [thick, quarter right, looseness=.8] (X) --[ thick, quarter right, looseness=.8] (x2r) ,
   (y1l) -- [thick, quarter left, looseness=.8] (Y), 
        (y2l) -- [thick, quarter right, looseness=.8] (Y),
   (Y)   -- [thick, quarter left, looseness=.8] (X1l)  ,
    (Y)   -- [thick, quarter right, looseness=.8] (Xnl)  ,
   (c1) -- [very thick, scalar, red] (c2),
};
\node[roundnode] (Y1) at (0+1.1+.1,0){\tiny{$\vspace{-.05cm}\hspace{-.0cm}\O_i$}};
\node[roundnode3] (X1) at (-.34,0){\tiny{$\vspace{-.05cm}\hspace{-.0cm}M$}};
  \end{feynman}
\end{tikzpicture}
  \end{minipage} 
  }

\newcommand{\twoloopgenTres}{
  \begin{minipage}[h]{0.12\linewidth}\begin{tikzpicture}
  [
  roundnode/.style={circle, draw=black!60, fill=black!30, very thick, 
  inner sep=2.1pt,
  text width=3mm},
roundnode3/.style={circle, draw=black!60, fill=black!6, very thick, 
  inner sep=2.5pt,
  text width=3mm},
]
\begin{feynman}[small]
\vertex (X) at (-.3,0);
\vertex (Xaux) at (-.3,-.1);
\vertex (x1l) at (-1.,.4);
\vertex (xnl) at (-1.,-.4);
\vertex (x1r) at (+.3,.3);
\vertex (x2r) at (+.3,-.3);
\vertex (c1) at (.4,.35);
\vertex (c2) at (.4,-.4);
\node [crossed dot] (Y) at (0+1.1+.1,0);
\vertex (Yaux) at (1.08+.1,-.09);
\vertex (y1l) at (-.5+1.,.3);
\vertex (y2l) at (-.5+1.,-.3);
\vertex (X1l) at (1.8+.1,.4);
\vertex (Xnl) at (1.8+.1,-.4);
   \vertex (t7) at (1.2,-.55){$\scriptsize{^{\text{1-loop}}}$};
   \diagram*{
   (x1l) -- [thick, quarter left, looseness=.8] (X) --[ thick, quarter left, looseness=.8] (x1r) ,
         (xnl) -- [thick, quarter right, looseness=.8] (X) --[ thick, quarter right, looseness=.8] (x2r) ,
   (y1l) -- [thick, quarter left, looseness=.8] (Y), 
        (y2l) -- [thick, quarter right, looseness=.8] (Y),
   (Y)   -- [thick, quarter left, looseness=.8] (X1l)  ,
    (Y)   -- [thick, quarter right, looseness=.8] (Xnl)  ,
   (c1) -- [very thick, scalar, red] (c2),
};
\node[roundnode] (Y1) at (0+1.1+.1,0){\tiny{$\vspace{-.05cm}\hspace{-.0cm}\O_i$}};
\node[roundnode3] (X1) at (-.34,0){\tiny{$\vspace{-.05cm}\hspace{-.0cm}M$}};
  \end{feynman}
\end{tikzpicture}
  \end{minipage} 
  }

\newcommand{\oneloopUn}{
  \begin{minipage}[h]{0.12\linewidth}\begin{tikzpicture}
  [
  roundnode/.style={circle, draw=black!60, fill=black!20, very thick, 
  inner sep=2.1pt,
  text width=3mm},
roundnode3/.style={circle, draw=black!60, fill=black!6, very thick, 
  inner sep=2.5pt,
  text width=3mm},
]
\begin{feynman}[small]
\vertex (X) at (-.3,0);
\vertex (Xaux) at (-.3,-.1);
\vertex (x1l) at (-1.,.4);
\vertex (xnl) at (-1.,-.4);
\vertex (x1r) at (+.3,.3);
\vertex (x2r) at (+.3,-.3);
\vertex (c1) at (.4,.35);
\vertex (c2) at (.4,-.4);
\node [crossed dot] (Y) at (0+1.1+.1,0);
\vertex (Yaux) at (1.08+.1,-.09);
\vertex (y1l) at (-.5+1.,.3);
\vertex (y2l) at (-.5+1.,-.3);
\vertex (X1l) at (1.8+.1,.4);
\vertex (Xnl) at (1.8+.1,-.4);
   \diagram*{
   (x1l) -- [thick, quarter left, looseness=.8] (X) --[ thick, quarter left, looseness=.8] (x1r) ,
         (xnl) -- [thick, quarter right, looseness=.8] (X) --[ thick, quarter right, looseness=.8] (x2r) ,
   (y1l) -- [thick, quarter left, looseness=.8] (Y), 
        (y2l) -- [thick, quarter right, looseness=.8] (Y),
   (Y)   -- [thick, quarter left, looseness=.8] (X1l)  ,
    (Y)   -- [thick, quarter right, looseness=.8] (Xnl)  ,
   (c1) -- [very thick, scalar, red] (c2),
};
\node[roundnode3] (Y1) at (0+1.1+.1,0){\tiny{$\vspace{-.05cm}\hspace{-.0cm}\O_{i}$}};
\node[roundnode3] (X1) at (-.34,0){\tiny{$\vspace{-.05cm}\hspace{-.0cm}M$}};
  \end{feynman}
\end{tikzpicture}
  \end{minipage} 
  }

\newcommand\TwoLoop{
  \begin{minipage}[h]{0.12\linewidth}\begin{tikzpicture}
  [
roundnode/.style={circle, draw=black!60, fill=black!6, very thick, 
  inner sep=1pt,
  text width=3mm},
]
\begin{feynman}[small]
\vertex   (X1) at (.1,-.2);
\vertex   (XX1) at (1.9,0-.2);
\vertex   (XXX1) at (2.1,0-.2);
\node [dot](X2) at (.75,0-.2) ;
\node [dot](X3) at (1.5,.8);
\vertex (y3) at (1.9,1.3);
\vertex (y4) at (1.9,.3);
\node [dot](X4) at (2.5,.8);
\vertex (y5) at (2.1,1.3);
\vertex (y6) at (2.1,.3);
\node [crossed dot, large] (X5) at (3.25,0-.2);
\vertex (X6) at (4,0-.2);
\vertex (X7) at (3.8,.8);
\node [dot] (X8) at (.8,1.1);
\vertex (X9) at (-.2,1.4);
\vertex (A1) at (1.8,1.5);
\vertex (A2) at (1.8, .5);
\vertex (A3) at (1.55, .2);
\vertex (A4) at (1.1, .3);
\vertex (A5) at (.5, .8);
\vertex (B1) at (1.8+.4,1.5);
\vertex (B2) at (1.8+.4, .5);
\vertex (B3) at (1.55+.4+2*.25, .2);
\vertex (B4) at (1.1+.4+2*.25+2*.45, .3);
\vertex (B5) at (.5+.4+2*.25+2*.45+2*.6, .8);
\vertex (C1) at (2,1.7);
\vertex (C2) at (2,-.6);
\vertex (t3) at (2.7,.9){$^\text{\tiny{$y_t$}}$};
\vertex (t3) at (1.3,.9){$^\text{\tiny{$y_t$}}$};
\vertex (t4) at (0.55,-.08){$^\text{\tiny{$y_\mu$}}$};
\vertex (t5) at (.9,1.3){$^\text{\tiny{$g^\prime$}}$};
\vertex (p2) at (-.3,1.4){$^\text{\tiny{$2$}}$};
\vertex (p1) at (0,-.24){$^\text{\tiny{$3$}}$};
\vertex (p4) at (3.85,.8){$^\text{\tiny{$4$}}$};
\vertex (p3) at (4.15,-.24){$^\text{\tiny{$1$}}$};
\vertex (p5) at (1.85,-.14){$^\text{\tiny{$x$}}$};
\vertex (p5) at (1.85,.46){$^\text{\tiny{$y$}}$};
\vertex (p5) at (1.85,1.37){$^\text{\tiny{$z$}}$};
\vertex (p5) at (3.7,-.43){$^\text{\tiny{$L_L^e$}}$};
\vertex (p5) at (3.8,+.43){$^\text{\tiny{$H$}}$};
\vertex (p5) at (2.7,-.43){$^\text{\tiny{$L_L^\mu$}}$};
\vertex (p5) at (.3,-.43){$^\text{\tiny{$e_R^\mu$}}$};
\vertex (p5) at (.3,1.4){$^\text{\tiny{$B$}}$};
   \diagram*{
 (X1)  -- [very thick] (X2) -- [very thick] (XX1) ,
 (XXX1) -- [very thick] (X5) -- [very thick, black] (X6),
 (X2)  -- [very thick, scalar, quarter left] (X3),
  (X4)  -- [very thick, scalar, quarter left] (X5),
  (X3) -- [very thick, quarter left] (y3),
  (y4) -- [very thick, quarter left] (X3),
  (X4) -- [very thick, quarter left] (y6),
  (y5) -- [very thick, quarter left] (X4),
  (X5) -- [very thick, scalar] (X7),
  (X8) -- [very thick, boson] (X9)
};
 \draw[-, to path={-| (\tikztotarget)}, very thick, rounded corners, draw=red, dashed] (C1) -- (C2);
  \end{feynman}
\end{tikzpicture}
  \end{minipage} 
  }

\newcommand\TwoLoopEXone{
  \begin{minipage}[h]{0.12\linewidth}\begin{tikzpicture}
  [
roundnode/.style={circle, draw=black!60, fill=black!6, very thick, 
  inner sep=1pt,
  text width=3mm},
]
\begin{feynman}[small]
\vertex   (X1) at (.1,-.2);
\vertex   (XX1) at (1.9,0-.2);
\node [dot](X2) at (.75,0-.2) ;
\node [dot](X3) at (1.5,.8);
\vertex (y3) at (1.9,1.3);
\vertex (y4) at (1.9,.3);
\node [dot] (X8) at (.8,1.1);
\vertex (X9) at (-.2,1.4);
   \diagram*{
 (X1)  -- [very thick] (X2) -- [very thick] (XX1) ,
 (X2)  -- [very thick, scalar, quarter left] (X3),
  (X3) -- [very thick, quarter left] (y3),
  (y4) -- [very thick, quarter left] (X3),
  (X8) -- [very thick, boson] (X9)
};
  \end{feynman}
\end{tikzpicture}
  \end{minipage} 
  }

\newcommand\TwoLoopEXtwo{
  \begin{minipage}[h]{0.12\linewidth}\begin{tikzpicture}
  [
roundnode/.style={circle, draw=black!60, fill=black!6, very thick, 
  inner sep=1pt,
  text width=3mm},
]
\begin{feynman}[small]
\vertex   (X1) at (.1,-.2);
\vertex   (XX1) at (1.9,0-.2);
\node [dot](X2) at (.75,0-.2) ;
\node [dot](X3) at (1.5,.8);
\vertex (y3) at (1.9,1.3);
\vertex (y4) at (1.9,.3);
\node [dot] (X8) at (1.4,-.2);
\vertex (X9) at (.5,-.7);
   \diagram*{
 (X1)  -- [very thick] (X2) -- [very thick] (XX1) ,
 (X2)  -- [very thick, scalar, quarter left] (X3),
  (X3) -- [very thick, quarter left] (y3),
  (y4) -- [very thick, quarter left] (X3),
  (X8) -- [very thick, boson] (X9)
};
  \end{feynman}
\end{tikzpicture}
  \end{minipage} 
  }

\newcommand\TwoLoopEXthree{
  \begin{minipage}[h]{0.12\linewidth}\begin{tikzpicture}
  [
roundnode/.style={circle, draw=black!60, fill=black!6, very thick, 
  inner sep=1pt,
  text width=3mm},
]
\begin{feynman}[small]
\vertex   (X1) at (.1,-.2);
\vertex   (XX1) at (1.9,0-.2);
\node [dot](X2) at (.75,0-.2) ;
\node [dot](X3) at (1.5,.8);
\vertex (y3) at (1.9,1.3);
\vertex (y4) at (1.9,.3);
\node [dot] (X8) at (.4,-.2);
\vertex (X9) at (-.5,-.7);
   \diagram*{
 (X1)  -- [very thick] (X2) -- [very thick] (XX1) ,
 (X2)  -- [very thick, scalar, quarter left] (X3),
  (X3) -- [very thick, quarter left] (y3),
  (y4) -- [very thick, quarter left] (X3),
  (X8) -- [very thick, boson] (X9)
};
  \end{feynman}
\end{tikzpicture}
  \end{minipage} 
  }

\newcommand\TwoLoopEXfour{
  \begin{minipage}[h]{0.12\linewidth}\begin{tikzpicture}
  [
roundnode/.style={circle, draw=black!60, fill=black!6, very thick, 
  inner sep=1pt,
  text width=3mm},
]
\begin{feynman}[small]
\vertex   (X1) at (.1,-.2);
\vertex   (XX1) at (1.9,0-.2);
\node [dot](X2) at (.75,0-.2) ;
\node [dot](X3) at (1.5,.8);
\vertex (y3) at (1.9,1.3);
\vertex (y4) at (1.9,.3);
\node [dot] (X8) at (1,+.6);
\vertex (X9) at (.1,+1.1);
   \diagram*{
 (X1)  -- [very thick] (X2) -- [very thick] (XX1) ,
 (X2)  -- [very thick, scalar, quarter left] (X3),
  (X3) -- [very thick, quarter left] (y3),
  (y4) -- [very thick, quarter left] (X3),
  (X8) -- [very thick, boson] (X9)
};
  \end{feynman}
\end{tikzpicture}
  \end{minipage} 
  }

\newcommand\TwoLoopEXfive{
  \begin{minipage}[h]{0.12\linewidth}\begin{tikzpicture}
  [
roundnode/.style={circle, draw=black!60, fill=black!6, very thick, 
  inner sep=1pt,
  text width=3mm},
]
\begin{feynman}[small]
\vertex   (X1) at (.1,-.2);
\vertex   (XX1) at (1.9,0-.2);
\node [dot](X2) at (.75,0-.2) ;
\node [dot](X3) at (1.5,.8);
\vertex (y3) at (1.9,1.3);
\vertex (y4) at (1.9,.3);
\node [dot] (X8) at (1.55,1.1);
\vertex (X9) at (.55,1.4);
\vertex (p5) at (1.85,.44){$^\text{\tiny{$t_L$}}$};
\vertex (p5) at (1.85,1.4){$^\text{\tiny{$t_R$}}$};
   \diagram*{
 (X1)  -- [very thick] (X2) -- [very thick] (XX1) ,
 (X2)  -- [very thick, scalar, quarter left] (X3),
  (X3) -- [very thick, quarter left] (y3),
  (y4) -- [very thick, quarter left] (X3),
  (X8) -- [very thick, boson] (X9)
};
  \end{feynman}
\end{tikzpicture}
  \end{minipage} 
  }

\newcommand\TwoLoopEXsix{
  \begin{minipage}[h]{0.12\linewidth}\begin{tikzpicture}
  [
roundnode/.style={circle, draw=black!60, fill=black!6, very thick, 
  inner sep=1pt,
  text width=3mm},
]
\begin{feynman}[small]
\vertex   (X1) at (.1,-.2);
\vertex   (XX1) at (1.9,0-.2);
\node [dot](X2) at (.75,0-.2) ;
\node [dot](X3) at (1.5,.8);
\vertex (y3) at (1.9,1.3);
\vertex (y4) at (1.9,.3);
\node [dot] (X8) at (1.55,1.1);
\vertex (X9) at (.55,1.4);
\vertex (p5) at (1.85,.44){$^\text{\tiny{$t_R$}}$};
\vertex (p5) at (1.85,1.4){$^\text{\tiny{$t_L$}}$};
   \diagram*{
 (X1)  -- [very thick] (X2) -- [very thick] (XX1) ,
 (X2)  -- [very thick, scalar, quarter left] (X3),
  (X3) -- [very thick, quarter left] (y3),
  (y4) -- [very thick, quarter left] (X3),
  (X8) -- [very thick, boson] (X9)
};
  \end{feynman}
\end{tikzpicture}
  \end{minipage} 
  }

\newcommand\TwoLoopR{
  \begin{minipage}[h]{0.12\linewidth}\begin{tikzpicture}
  [
roundnode/.style={circle, draw=black!60, fill=black!6, very thick, 
  inner sep=1pt,
  text width=3mm},
]
\begin{feynman}[small]
\vertex   (X1) at (.1,-.2);
\vertex   (XX1) at (1.9,0-.2);
\vertex   (XXX1) at (2.1,0-.2);
\node [dot](X2) at (.75,0-.2) ;
\node [dot](X3) at (1.5,.8);
\vertex (y3) at (1.9,1.3);
\vertex (y4) at (1.9,.3);
\node [dot](X4) at (2.5,.8);
\vertex (y5) at (2.1,1.3);
\vertex (y6) at (2.1,.3);
\node [crossed dot, large] (X5) at (3.25,0-.2);
\vertex (X6) at (4,0-.2);
\vertex (X7) at (3.8,.8);
\node [dot] (X8) at (.8,1.1);
\vertex (X9) at (-.2,1.4);
\vertex (A1) at (1.8,1.5);
\vertex (A2) at (1.8, .5);
\vertex (A3) at (1.55, .2);
\vertex (A4) at (1.1, .3);
\vertex (A5) at (.5, .8);
\vertex (B1) at (1.8+.4,1.5);
\vertex (B2) at (1.8+.4, .5);
\vertex (B3) at (1.55+.4+2*.25, .2);
\vertex (B4) at (1.1+.4+2*.25+2*.45, .3);
\vertex (B5) at (.5+.4+2*.25+2*.45+2*.6, .8);
\vertex (C1) at (2,1.7);
\vertex (C2) at (2,-.6);
\vertex (t3) at (2.7,.9){$^\text{\tiny{$y_t$}}$};
\vertex (t3) at (1.3,.9){$^\text{\tiny{$y_t$}}$};
\vertex (t4) at (0.55,-.08){$^\text{\tiny{$y_e$}}$};
\vertex (t5) at (.9,1.3){$^\text{\tiny{$g^\prime$}}$};
\vertex (p2) at (-.3,1.4){$^\text{\tiny{$2$}}$};
\vertex (p1) at (0,-.24){$^\text{\tiny{$1$}}$};
\vertex (p4) at (3.85,.8){$^\text{\tiny{$4$}}$};
\vertex (p3) at (4.15,-.24){$^\text{\tiny{$3$}}$};
\vertex (p5) at (1.85,-.14){$^\text{\tiny{$x$}}$};
\vertex (p5) at (1.85,.46){$^\text{\tiny{$y$}}$};
\vertex (p5) at (1.85,1.37){$^\text{\tiny{$z$}}$};
\vertex (p5) at (3.7,-.43){$^\text{\tiny{$e_R^\mu$}}$};
\vertex (p5) at (3.8,+.43){$^\text{\tiny{$H$}}$};
\vertex (p5) at (2.7,-.43){$^\text{\tiny{$e_R^e$}}$};
\vertex (p5) at (.3,-.43){$^\text{\tiny{$L_L^e$}}$};
\vertex (p5) at (.3,1.4){$^\text{\tiny{$B$}}$};
   \diagram*{
 (X1)  -- [very thick] (X2) -- [very thick] (XX1) ,
 (XXX1) -- [very thick] (X5) -- [very thick, black] (X6),
 (X2)  -- [very thick, scalar, quarter left] (X3),
  (X4)  -- [very thick, scalar, quarter left] (X5),
  (X3) -- [very thick, quarter left] (y3),
  (y4) -- [very thick, quarter left] (X3),
  (X4) -- [very thick, quarter left] (y6),
  (y5) -- [very thick, quarter left] (X4),
  (X5) -- [very thick, scalar] (X7),
  (X8) -- [very thick, boson] (X9)
};
 \draw[-, to path={-| (\tikztotarget)}, very thick, rounded corners, draw=red, dashed] (C1) -- (C2);
  \end{feynman}
\end{tikzpicture}
  \end{minipage} 
  }

\newcommand\OneLoop{
  \begin{minipage}[h]{0.12\linewidth}\begin{tikzpicture}
  [
roundnode/.style={circle, draw=black!60, fill=black!6, very thick, 
  inner sep=1pt,
  text width=3mm},
]
\begin{feynman}[small]
\vertex   (X1) at (.1-1.2,-.2);
\vertex   (XX1) at (3.1-1.2,0-.2);
\node [dot](X2) at (.75-1.,0-.2) ;
\node [dot](X3) at (1.5-1.,.6);
\node [dot](X4) at (2.5-1.2,.6);
\vertex(X44) at (3.1-1.2,.6);
\node [dot] (X8) at (.8-1.2,.7);
\vertex (X9) at (-.2-1.2,1.);
\vertex   (XXX1) at (2.1,0-.2);
\vertex (X444) at (3.1-1,.6);
\node [crossed dot, large] (X5) at (3.,0-.2);
\vertex (X6) at (4,0-.2);
\vertex (X7) at (3.8,.6);
\vertex (A1) at (1.8,1.5);
\vertex (A2) at (1.8, .5);
\vertex (A3) at (1.55, .2);
\vertex (A4) at (1.1, .3);
\vertex (A5) at (.5, .6);
\vertex (B1) at (1.8+.4,1.5);
\vertex (B2) at (1.8+.4, .5);
\vertex (B3) at (1.55+.4+2*.25, .2);
\vertex (B4) at (1.1+.4+2*.25+2*.45, .3);
\vertex (B5) at (.5+.4+2*.25+2*.45+2*.6, .8);
\vertex (C1) at (2,1.3);
\vertex (C2) at (2,-.55);
\vertex (t1) at (-.9,-.45){$^\text{\tiny{$3$}}$};
\vertex (t2) at (1.8,-.45){$^\text{\tiny{$x$}}$};
\vertex (t2) at (1.8,.68){$^\text{\tiny{$y$}}$};
\vertex (t2) at (-1.1,1.07){$^\text{\tiny{$2$}}$};
   \diagram*{
 (X1)  -- [very thick] (X2) -- [very thick] (XX1) ,
 (XXX1) -- [very thick] (X5) -- [very thick, black] (X6),
 (X2)  -- [very thick, scalar, quarter left] (X3),
  (X4)  -- [very thick, scalar ] (X44),
  (X3) --  [very thick, half left] (X4)  --  [very thick, half left] (X3),
    (X444)  -- [very thick, scalar , quarter left, looseness=1] (X5),
  (X5) -- [very thick, scalar] (X7),
  (X8) -- [very thick, boson] (X9)
};
 \draw[-, to path={-| (\tikztotarget)}, very thick, rounded corners, draw=red, dashed] (C1) -- (C2);
  \end{feynman}
\end{tikzpicture}
  \end{minipage} 
  }
  
\newcommand\QuarticOne{
  \begin{minipage}[h]{0.12\linewidth}\begin{tikzpicture}
  [
roundnode/.style={circle, draw=black!60, fill=black!6, very thick, 
  inner sep=1pt,
  text width=3mm},
]
\begin{feynman}[small]

\vertex   (XXX1) at (2.1,0-.2);
\vertex (X2) at (.18,.55) ;
\node [dot](X3) at (1.1,.55);
\vertex (X33) at (1.9,.55);
\vertex (y3) at (1.9,1.3);
\vertex (y4) at (1.9,-.2);
\node [crossed dot, large] (X4) at (2.9,.55);
\vertex (y5) at (2.1,1.3);
\vertex (y6) at (2.1,.55);
\node [dot] (X5) at (2.9,-.2);
\vertex (X6) at (3.9,0-.2);
\vertex (X7) at (3.9,.55);
\node [dot] (X8) at (.8,1.1);
\vertex (X9) at (-.2,1.4);
\vertex (A1) at (1.8,1.5);
\vertex (A2) at (1.8, .5);
\vertex (A3) at (1.55, .2);
\vertex (A4) at (1.1, .3);
\vertex (A5) at (.5, .8);
\vertex (B1) at (1.8+.4,1.5);
\vertex (B2) at (1.8+.4, .5);
\vertex (B3) at (1.55+.4+2*.25, .2);
\vertex (B4) at (1.1+.4+2*.25+2*.45, .3);
\vertex (B5) at (.5+.4+2*.25+2*.45+2*.6, .8);
\vertex (C1) at (2,1.7);
\vertex (C2) at (2,-.6);
\vertex (t3) at (.95,.65){$^\text{\tiny{$\lambda$}}$};
\vertex (t3) at (3.15,-.06){$^\text{\tiny{$y_\mu$}}$};
\vertex (p2) at (-.3,1.4){$^\text{\tiny{$2$}}$};
\vertex (p1) at (.05,.5){$^\text{\tiny{$4$}}$};
\vertex (p4) at (4.05,.5){$^\text{\tiny{$1$}}$};
\vertex (p3) at (4.05,-.24){$^\text{\tiny{$3$}}$};
\vertex (p5) at (1.85,-.14){$^\text{\tiny{$x$}}$};
\vertex (p5) at (1.85,.65){$^\text{\tiny{$y$}}$};
\vertex (p5) at (1.85,1.37){$^\text{\tiny{$z$}}$};
\vertex (p5) at (3.5,-.43){$^\text{\tiny{$e_R^\mu$}}$};
\vertex (p5) at (.4,+.32){$^\text{\tiny{$H$}}$};
\vertex (p5) at (3.5,.7){$^\text{\tiny{$L_L^e$}}$};
\vertex (p5) at (.3,1.4){$^\text{\tiny{$B$}}$};
   \diagram*{
 (X2)  -- [very thick, scalar] (X3) -- [very thick, scalar] (X33),
 (X3) -- [very thick, quarter left, scalar] (y3),
  (y4) -- [very thick, quarter left, scalar] (X3),
 (XXX1) -- [very thick,scalar] (X5) -- [very thick, black] (X6),
  (X4)  -- [very thick ] (X5),
  (X4) -- [very thick, scalar] (y6),
  (y5) -- [very thick, quarter left, scalar] (X4),
  (X4) -- [very thick] (X7),
  (X8) -- [very thick, boson] (X9)
};
 \draw[-, to path={-| (\tikztotarget)}, very thick, rounded corners, draw=red, dashed] (C1) -- (C2);
  \end{feynman}
\end{tikzpicture}
  \end{minipage} 
  }

\newcommand\QuarticTwo{
  \begin{minipage}[h]{0.12\linewidth}\begin{tikzpicture}
  [
roundnode/.style={circle, draw=black!60, fill=black!6, very thick, 
  inner sep=1pt,
  text width=3mm},
]
\begin{feynman}[small]

\vertex   (XXX1) at (2.1,0-.2);
\vertex (X2) at (.18,.55) ;
\node [dot](X3) at (1.1,.55);
\vertex (X33) at (1.9,.55);
\vertex (y3) at (1.9,1.3);
\node [dot] (y33) at (1.1,-.2);
\vertex (y333) at (.18,-.2);
\vertex (y4) at (1.9,-.2);
\node [crossed dot, large] (X4) at (2.9,.55);
\vertex (y5) at (2.1,1.3);
\vertex (y6) at (2.1,.55);
\vertex (X6) at (3.9,0-.2);
\vertex (X7) at (3.9,.55);
\node [dot] (X8) at (3.2,1.3);
\vertex (X9) at (4.2,1.6);
\vertex (A1) at (1.8,1.5);
\vertex (A2) at (1.8, .5);
\vertex (A3) at (1.55, .2);
\vertex (A4) at (1.1, .3);
\vertex (A5) at (.5, .8);
\vertex (B1) at (1.8+.4,1.5);
\vertex (B2) at (1.8+.4, .5);
\vertex (B3) at (1.55+.4+2*.25, .2);
\vertex (B4) at (1.1+.4+2*.25+2*.45, .3);
\vertex (B5) at (.5+.4+2*.25+2*.45+2*.6, .8);
\vertex (C1) at (2,1.7);
\vertex (C2) at (2,-.6);
\vertex (t3) at (.95,.65){$^\text{\tiny{$\lambda$}}$};
\vertex (t3) at (.9,-.06){$^\text{\tiny{$y_\mu$}}$};
\vertex (p2) at (4.35,1.55){$^\text{\tiny{$2$}}$};
\vertex (p1) at (.05,.5){$^\text{\tiny{$4$}}$};
\vertex (p4) at (4.05,.5){$^\text{\tiny{$1$}}$};
\vertex (p3) at (.05,-.24){$^\text{\tiny{$3$}}$};
\vertex (p5) at (1.85,-.14){$^\text{\tiny{$x$}}$};
\vertex (p5) at (1.85,.65){$^\text{\tiny{$y$}}$};
\vertex (p5) at (1.85,1.37){$^\text{\tiny{$z$}}$};
\vertex (p5) at (.5,-.43){$^\text{\tiny{$e_R^\mu$}}$};
\vertex (p5) at (.4,+.32){$^\text{\tiny{$H$}}$};
\vertex (p5) at (3.5,.7){$^\text{\tiny{$L_L^e$}}$};
\vertex (p5) at (3.6,1.57){$^\text{\tiny{$B$}}$};
   \diagram*{
 (X2)  -- [very thick, scalar] (X3) -- [very thick, scalar] (X33),
 (X3) -- [very thick, quarter left, scalar] (y3),
  (y33) -- [very thick, scalar] (X3),
  (y333)--  [very thick] (y33) --  [very thick ] (y4), 
 (XXX1) -- [very thick, quarter right ] (X4),
  (X4) -- [very thick, scalar] (y6),
  (y5) -- [very thick, quarter left, scalar] (X4),
  (X4) -- [very thick] (X7),
  (X8) -- [very thick, boson] (X9)
};
 \draw[-, to path={-| (\tikztotarget)}, very thick, rounded corners, draw=red, dashed] (C1) -- (C2);
  \end{feynman}
\end{tikzpicture}
  \end{minipage} 
  }

\newcommand\QuarticThree{
  \begin{minipage}[h]{0.12\linewidth}\begin{tikzpicture}
  [
roundnode/.style={circle, draw=black!60, fill=black!6, very thick, 
  inner sep=1pt,
  text width=3mm},
]
\begin{feynman}[small]

\vertex   (XXX1) at (2.1,0-.2);
\vertex (X2) at (.18,.55) ;
\node [dot](X3) at (1.1,.55);
\vertex (X33) at (1.9,.55);
\vertex (y3) at (1.9,1.3);
\vertex (y4) at (1.9,-.2);
\node [crossed dot, large] (X4) at (2.9,.55);
\vertex (y5) at (2.1,1.3);
\vertex (y6) at (2.1,.55);
\node [dot] (X5) at (2.9,-.2);
\vertex (X6) at (3.9,0-.2);
\vertex (X7) at (3.9,.55);
\node [dot] (X8) at (.8,1.1);
\vertex (X9) at (-.2,1.4);
\vertex (A1) at (1.8,1.5);
\vertex (A2) at (1.8, .5);
\vertex (A3) at (1.55, .2);
\vertex (A4) at (1.1, .3);
\vertex (A5) at (.5, .8);
\vertex (B1) at (1.8+.4,1.5);
\vertex (B2) at (1.8+.4, .5);
\vertex (B3) at (1.55+.4+2*.25, .2);
\vertex (B4) at (1.1+.4+2*.25+2*.45, .3);
\vertex (B5) at (.5+.4+2*.25+2*.45+2*.6, .8);
\vertex (C1) at (2,1.7);
\vertex (C2) at (2,-.6);
\vertex (t3) at (.95,.65){$^\text{\tiny{$\lambda$}}$};
\vertex (t3) at (3.15,-.06){$^\text{\tiny{$y_e$}}$};
\vertex (p2) at (-.3,1.4){$^\text{\tiny{$2$}}$};
\vertex (p1) at (.05,.5){$^\text{\tiny{$4$}}$};
\vertex (p4) at (4.05,.5){$^\text{\tiny{$3$}}$};
\vertex (p3) at (4.05,-.24){$^\text{\tiny{$1$}}$};
\vertex (p5) at (1.85,-.14){$^\text{\tiny{$x$}}$};
\vertex (p5) at (1.85,.65){$^\text{\tiny{$y$}}$};
\vertex (p5) at (1.85,1.37){$^\text{\tiny{$z$}}$};
\vertex (p5) at (3.5,-.43){$^\text{\tiny{$L_L^e$}}$};
\vertex (p5) at (.4,+.32){$^\text{\tiny{$H$}}$};
\vertex (p5) at (3.5,.7){$^\text{\tiny{$e_R^\mu$}}$};
\vertex (p5) at (.3,1.4){$^\text{\tiny{$B$}}$};
   \diagram*{
 (X2)  -- [very thick, scalar] (X3) -- [very thick, scalar] (X33),
 (X3) -- [very thick, quarter left, scalar] (y3),
  (y4) -- [very thick, quarter left, scalar] (X3),
 (XXX1) -- [very thick,scalar] (X5) -- [very thick, black] (X6),
  (X4)  -- [very thick ] (X5),
  (X4) -- [very thick, scalar] (y6),
  (y5) -- [very thick, quarter left, scalar] (X4),
  (X4) -- [very thick] (X7),
  (X8) -- [very thick, boson] (X9)
};
 \draw[-, to path={-| (\tikztotarget)}, very thick, rounded corners, draw=red, dashed] (C1) -- (C2);
  \end{feynman}
\end{tikzpicture}
  \end{minipage} 
  }

\newcommand\QuarticFour{
  \begin{minipage}[h]{0.12\linewidth}\begin{tikzpicture}
  [
roundnode/.style={circle, draw=black!60, fill=black!6, very thick, 
  inner sep=1pt,
  text width=3mm},
]
\begin{feynman}[small]

\vertex   (XXX1) at (2.1,0-.2);
\vertex (X2) at (.18,.55) ;
\node [dot](X3) at (1.1,.55);
\vertex (X33) at (1.9,.55);
\vertex (y3) at (1.9,1.3);
\node [dot] (y33) at (1.1,-.2);
\vertex (y333) at (.18,-.2);
\vertex (y4) at (1.9,-.2);
\node [crossed dot, large] (X4) at (2.9,.55);
\vertex (y5) at (2.1,1.3);
\vertex (y6) at (2.1,.55);
\vertex (X6) at (3.9,0-.2);
\vertex (X7) at (3.9,.55);
\node [dot] (X8) at (3.2,1.3);
\vertex (X9) at (4.2,1.6);
\vertex (A1) at (1.8,1.5);
\vertex (A2) at (1.8, .5);
\vertex (A3) at (1.55, .2);
\vertex (A4) at (1.1, .3);
\vertex (A5) at (.5, .8);
\vertex (B1) at (1.8+.4,1.5);
\vertex (B2) at (1.8+.4, .5);
\vertex (B3) at (1.55+.4+2*.25, .2);
\vertex (B4) at (1.1+.4+2*.25+2*.45, .3);
\vertex (B5) at (.5+.4+2*.25+2*.45+2*.6, .8);
\vertex (C1) at (2,1.7);
\vertex (C2) at (2,-.6);
\vertex (t3) at (.95,.65){$^\text{\tiny{$\lambda$}}$};
\vertex (t3) at (.9,-.06){$^\text{\tiny{$y_e$}}$};
\vertex (p2) at (4.35,1.55){$^\text{\tiny{$2$}}$};
\vertex (p1) at (.05,.5){$^\text{\tiny{$4$}}$};
\vertex (p4) at (4.05,.5){$^\text{\tiny{$3$}}$};
\vertex (p3) at (.05,-.24){$^\text{\tiny{$1$}}$};
\vertex (p5) at (1.85,-.14){$^\text{\tiny{$x$}}$};
\vertex (p5) at (1.85,.65){$^\text{\tiny{$y$}}$};
\vertex (p5) at (1.85,1.37){$^\text{\tiny{$z$}}$};
\vertex (p5) at (.5,-.43){$^\text{\tiny{$L_L^e$}}$};
\vertex (p5) at (.4,+.32){$^\text{\tiny{$H$}}$};
\vertex (p5) at (3.5,.7){$^\text{\tiny{$e_R^\mu$}}$};
\vertex (p5) at (3.6,1.57){$^\text{\tiny{$B$}}$};
   \diagram*{
 (X2)  -- [very thick, scalar] (X3) -- [very thick, scalar] (X33),
 (X3) -- [very thick, quarter left, scalar] (y3),
  (y33) -- [very thick, scalar] (X3),
  (y333)--  [very thick] (y33) --  [very thick ] (y4), 
 (XXX1) -- [very thick, quarter right ] (X4),
  (X4) -- [very thick, scalar] (y6),
  (y5) -- [very thick, quarter left, scalar] (X4),
  (X4) -- [very thick] (X7),
  (X8) -- [very thick, boson] (X9)
};
 \draw[-, to path={-| (\tikztotarget)}, very thick, rounded corners, draw=red, dashed] (C1) -- (C2);
  \end{feynman}
\end{tikzpicture}
  \end{minipage} 
  }

\newcommand\OneLoopFirst{
  \begin{minipage}[h]{0.12\linewidth}\begin{tikzpicture}
  [
roundnode/.style={circle, draw=black!60, fill=black!6, very thick, 
  inner sep=1pt,
  text width=3mm},
]
\begin{feynman}[small]
\vertex (X2) at (.6,.8) ;
\node [dot](X3) at (1.5,.8);
\vertex (y3) at (1.9,1.3);
\vertex (y4) at (1.9,.3);
\node [crossed dot, large] (X4) at (2.5,.8);
\vertex (y5) at (2.1,1.3);
\vertex (y6) at (2.1,.3);
\vertex (y7) at (3.3,1.3);
\vertex (y8) at (3.3,.3);
\node [dot] (X8) at (1,1.3);
\vertex (X9) at (.1,1.7);
\vertex (C1) at (2,1.7);
\vertex (C2) at (2,-.1);
\vertex (t3) at (1.3,.9){$^\text{\tiny{$y_t$}}$};
\vertex (t5) at (1.25,1.3){$^\text{\tiny{$g^\prime$}}$};
\vertex (p2) at (0,1.7){$^\text{\tiny{$2$}}$};
\vertex (p4) at (.47,.72){$^\text{\tiny{$4$}}$};
\vertex (p3) at (3.6,.24){$^\text{\tiny{$1_L^-$}}$};
\vertex (p3) at (3.6,1.35){$^\text{\tiny{$3^-$}}$};
\vertex (p5) at (1.85,.46){$^\text{\tiny{$y$}}$};
\vertex (p5) at (1.8,1.37){$^\text{\tiny{$x_a$}}$};
\vertex (p5) at (3.,.0){$^\text{\tiny{$L_L^e$}}$};
\vertex (p5) at (.8,.55){$^\text{\tiny{$H$}}$};
\vertex (p5) at (3,1.55){$^\text{\tiny{$e_R^\mu$}}$};
\vertex (p5) at (.6,1.65){$^\text{\tiny{$B$}}$};
   \diagram*{
 (X2)  -- [very thick, scalar] (X3),
  (X3) -- [very thick, quarter left] (y3),
  (y4) -- [very thick, quarter left] (X3),
  (X4) -- [very thick, quarter left] (y6),
  (y5) -- [very thick, quarter left] (X4),
  (y7) -- [very thick, quarter right] (X4),
  (y8) -- [very thick, quarter left] (X4),
  (X8) -- [very thick, boson] (X9)
};
 \draw[-, to path={-| (\tikztotarget)}, very thick, rounded corners, draw=red, dashed] (C1) -- (C2);
  \end{feynman}
\end{tikzpicture}
  \end{minipage} 
  }

\begin{document}

\begin{titlepage}
\begin{center}
\hfill

\vspace{2.0cm}
{\Large\bf  
Gearing up for the next generation of LFV \\[.5cm]  experiments, via on-shell methods 
}
\vspace{2.0cm}
\\
{\bf 
Joan Elias Mir\'o$^{a}$,
Clara Fernandez$^{d}$,
Mehmet Asım Gümüş$^{b,c}$
 and
Alex Pomarol$^{d,e}$
}
\\
\vspace{0.7cm}
{\it\footnotesize
${}^a$ICTP, International Centre for Theoretical Physics, strada Costiera 11, 34151, Trieste, Italy  \\
\vspace{0.1cm}
${}^b$SISSA, Via Bonomea 265, I-34136 Trieste, Italy \\[-.1cm]
\vspace{0.1cm}
${}^c$INFN, Sezione di Trieste, Via Valerio 2, 34127 Trieste, Italy \\
\vspace{0.1cm}
${}^d$IFAE and BIST, Universitat Aut\`onoma de Barcelona, 08193 Bellaterra, Barcelona \\[-.1cm]
\vspace{0.1cm}
${}^e$Departament de F\'isica, Universitat Aut\`onoma de Barcelona, 08193 Bellaterra, Barcelona
}

\vspace{0.9cm}
\abstract

Lepton Flavor Violating (LFV) observables such as $\mu\to e\gamma$, $\mu\to 3e$ and $\mu N \to eN$
are among the best probes for  new physics at the TeV scale. In the near future the bounds on these observables will 
improve by many orders of magnitude.
In this work we use the SM EFT to understand the impact of these measurements. 
The precision  reach 
 is such that the interpretation  of the bounds requires an analysis of the  dimension-six operator mixing up to  the two-loop level.
Using on-shell amplitude techniques, which  make transparent many selection rules, 
we   classify and calculate the different operator mixing chains.
At the leading order, on-shell techniques allow to calculate anomalous dimensions of SM EFT operators  from  the product of tree-level amplitudes, even for two-loop renormalization group mixings.
We illustrate the importance of our EFT  approach  in    models with extra vector-like fermions. 

\end{center}
\end{titlepage}
\setcounter{footnote}{0}

\setcounter{tocdepth}{1} 
\tableofcontents


\section{Introduction}

In the SM, lepton numbers $L_{e,\mu,\tau}$ arise from accidental symmetries that guarantee the absence of neutrino masses and lepton flavor violating (LFV) processes. These symmetries are, however,  not respected by  higher-dimensional operators beyond the SM   suppressed by some new physics scale $\Lambda$. In fact, already at the first order in the $1/\Lambda$ expansion (dimension-five operators), we find that  lepton number can be broken in two units, $\Delta L_i=2$,  generating then neutrino masses.  The smallness of the neutrino masses however forces  the scale  suppressing these operators $\Lambda$ to be extremely large (or the corresponding Wilson coefficients extremely small), 
making the physical effects  of these operators only visible in neutrino physics.

We are here interested  in LFV processes where the relative lepton numbers  are violated, but
the total lepton number is preserved.
These processes are among the best indirect probes for new physics at the TeV~\cite{Calibbi:2017uvl}.
This is due to the fact that  $L_{e,\mu,\tau}$ are not  easy to arise accidentally 
 in scenarios beyond the SM (BSM), where the  matter content and interactions are much  
more extended than the simple leptonic sector of the SM.
As a consequence, sizable  BSM effects are expected to arise in  LFV processes.
They can be characterized  by  dimension-six operators  ($\sim 1/\Lambda^2$) that, 
as they preserve the total lepton number,
do  not need to be as highly suppressed as dimension-five operators.
For this reason they can  have an important impact in future LFV experiments involving the charged lepton sector.

We will consider in particular LFV processes with $\Delta L_e=\Delta L_\mu=1$.
At present,    
the most competitive   experimental measurements come from
the processes $\mu\to e\gamma$,  $\mu\to eee$  and $\mu N\to e N$  
(see Table~\ref{tabbb}), and these  will  improve in the near future.
Especially, the sensitivities of $\mu\to eee$  and $\mu N\to e N$ are expected to improve
 by  four orders of magnitude by the mid-2020s~\cite{Baldini:2018uhj}.
These LFV processes  are very clean observables from the theoretical point of view, since the contributions of the SM (even when implemented with neutrino masses)  are much smaller than the present and future experimental sensitivities.
We will also consider 
$h\to \mu  e$, 
although, as we will show, it does not place  any important constraint.
There are other LFV processes such as $Z\to \mu  e$ and $J/\psi\to \mu e$,
but they are not competitive  with those in Table~\ref{tabbb}.
We also have processes involving  flavor violations in the quark sector, e.g. $K\to  \mu e$, 
but these quark flavor transitions
 are quite constrained from other non LFV processes,  so we will not consider them here.

We will use the Effective Field Theory (EFT) approach to characterize the 
BSM contributions to  LFV processes   as they are captured in the Wilson coefficients of  the dimension-six operators.
Our purpose is to understand at what order in the loop expansion the  Wilson coefficients can enter into the different observables
of Table~\ref{tabbb}.
Depending on the experimental sensitivity, we will see that certain Wilson coefficients can be better bounded
by $\mu\to e\gamma$ even when  they enter at the  two-loop level.

This program will therefore require the calculation of  anomalous dimensions at the two-loop level.
We will use on-shell methods to perform these calculations. It has already been shown 
the efficiency of these methods to calculate anomalous dimensions at the loop level~\cite{Caron-Huot:2016cwu,EliasMiro:2020tdv,Baratella:2020lzz,Jiang:2020mhe,Bern:2020ikv,Baratella:2021guc,AccettulliHuber:2021uoa},
where  these can be  reduced to  products of  tree-level amplitudes integrated over a phase space.
On-shell amplitude methods are also able to show in a transparent way 
many selection rules hidden in the ordinary Feynman approach 
\cite{Elias-Miro:2014eia,Cheung:2015aba,Bern:2019wie,Craig:2019wmo,Jiang:2020rwz,Baratella:2020dvw,
Li:2020zfq,Shu:2021qlr}, 
mainly due to helicity~\cite{Cheung:2015aba} or angular momentum conservation~\cite{Jiang:2020rwz,Baratella:2020dvw}. This    simplifies  substantially the loop calculations.\footnote{On-shell techniques for  the SM EFT have also  been applied in related contexts~\cite{Shadmi:2018xan,Durieux:2019eor,Durieux:2019siw,Dong:2021yak}.}

Our calculation of the anomalous dimensions at the two-loop level will
in particular show  how  $\mu\to e\gamma$
can constrain the LFV couplings  $W\mu \nu_e$ and $h\mu e$ 
and some four-fermion interactions  at an unprecedented level.
Previous (partial) analysis can be found in Refs.~\cite{Crivellin:2013hpa,Pruna:2014asa,Pruna:2015jhf,Crivellin:2017rmk,Ardu:2021koz,Davidson:2020hkf,Davidson:2016edt}.\footnote{For similar studies on LFV processes involving $\tau$ leptons, see for example Refs.~\cite{Celis:2013xja,Celis:2014asa, Husek:2020fru, Cirigliano:2021img}.}

In Section~\ref{sec2} we review the LFV  dimension-six operators of the SM EFT  
and the tree-level bounds derived from the processes in  Table~\ref{tabbb}. 
In Section~\ref{muegz} we analyze which operators  mix into the dipole operators responsible for $\mu\to e\gamma$ .  
We find that there are various  important  two-loop anomalous dimension matrix elements that are unknown. 
Section~\ref{twoloopd} is then dedicated to compute these missing pieces 
of the  two-loop renormalization group (RG) equation, via on-shell methods. 
In Sections \ref{mueg} and \ref{muNeN} we analyze the  RG mixing of dimension-six operators into the $\mu\rightarrow eee$ and $\mu N \rightarrow e N$ processes, respectively. 
A global perspective  of the new loop effects we have found from  mixings into the LFV observables is summarized in Section~\ref{constADM}.
For illustration,  in Section~\ref{models} we show the impact of our analysis on two simple UV completions of the SM EFT.
 Finally, we conclude in Section~\ref{conclu}.  
 
  { \renewcommand{\arraystretch}{1.3} \renewcommand\tabcolsep{6.4pt}
\begin{table}[t]
\begin{center}
\begin{tabular}{L{2.5cm} L{2.5cm}  L{2.5cm}  L{2.5cm}  L{2.5cm}   L{2.5cm}   }  
\toprule  
  & BR($\mu \to e \gamma $) & BR($\mu \to eee$)  & $R(\mu N \to e N)$  & BR($h \to \mu e $) 
     \\
   \rowcolor{Gray1}
\midrule
      \bf{Current}    &  $4.2\cdot 10^{-13}$~\cite{MEG:2016leq}  &     $1\cdot 10^{-12}$~\cite{SINDRUM:1987nra} &     $7\cdot 10^{-13}$~\cite{SINDRUMII:2006dvw}   &   $6.1\cdot 10^{-5}$~\cite{ATLAS:2019old}     
       \\  
 \bf{Future}              &   $6.0 \cdot 10^{-14}$~\cite{MEGII:2018kmf}  &    $1\cdot 10^{-16}$~\cite{Blondel:2013ia}   & $8\cdot 10^{-17}$~\cite{Mu2e:2014fns} \\   
 
   \bottomrule
\end{tabular}
\caption[Works]{ 
Current and near future upper bounds  on   $\Delta L_e=\Delta L_\mu=1$ processes.     \label{tabbb} }
\end{center} \vspace{-.5cm}
\end{table}
}

\section{LFV dimension-six operator basis}
\label{sec2}

The relevant dimension-six operators for our analysis 
of  processes with $\Delta L_e=\Delta L_\mu=1$ 
are given by
\bea
{\cal L}_{6}&=& C^{\mu e}_{DW}\frac{ y_\mu g}{ \Lambda^2}\bar{L}^{(2)}_L\tau^a\sigma^{\mu\nu} e^{(1)}_R HW^a_{\mu\nu} +C^{\mu e}_{DB} \frac{ y_\mu g^\prime}{ \Lambda^2}\bar{L}^{(2)}_L\sigma^{\mu\nu} e^{(1)}_RHB_{\mu\nu} 
+(\mu\leftrightarrow e)
 \label{opD}\\[.2cm]
&+& 
\frac{C^{\mu e}_{L} }{\Lambda^2}(H^\dagger i \overset{\leftrightarrow}{D}_\mu H)(\bar{L}^{(2)}_L\gamma^\mu L^{(1)}_L)
 + 
\frac{C^{\mu e}_{L3}}{\Lambda^2}(H^\dagger i \overset{\leftrightarrow}{D^a_\mu} H)(\bar{L}^{(2)}_L\tau^a\gamma^\mu L^{(1)}_L)
+ 
\frac{C^{\mu e}_{R}}{\Lambda^2}(H^\dagger i \overset{\leftrightarrow}{D}_\mu H)(\bar{e}^{(2)}_R\gamma^\mu e^{(1)}_R) 
   \  \ \ \ \ \ \ \  \label{opJJ}\\[.2cm]
&+&
 C^{\mu e}_{y}\frac{  y_\mu }{\Lambda^2}\left(H^\dagger H\right)\left(\bar{L}^{(2)}_Le^{(1)}_R H\right) 
 +(\mu\leftrightarrow e)
   \label{opy}\\[.2cm]
&+& 
 \frac{C^{\mu eff}_{LL}}{\Lambda^2} (\bar{L}^{(2)}_L \gamma_\mu L^{(1)}_L)(\bar{F}_L \gamma^\mu F_L)
 + \frac{C^{\mu eff}_{LL3}}{\Lambda^2} (\bar{L}^{(2)}_L \tau^a\gamma_\mu L^{(1)}_L)(\bar{F}_L \tau^a \gamma^\mu F_L)
+ \frac{C^{\mu eff}_{RR}}{\Lambda^2} (\bar{e}^{(2)}_R \gamma_\mu e^{(1)}_R)(\bar{f}_R \gamma^\mu f_R)
 \nonumber \\[.2cm]
&+& 
 \frac{C^{\mu eff}_{LR}}{\Lambda^2} (\bar{L}^{(2)}_L \gamma_\mu L^{(1)}_L)(\bar{f}_R \gamma^\mu f_R)
+ \frac{C^{\mu eff}_{RL}}{\Lambda^2} (\bar{e}^{(2)}_R \gamma^\mu e^{(1)}_R)(\bar{F}_L \gamma_\mu F_L)
 \nonumber \\[.2cm]
&+& 
C^{\mu lle}_{LR}\frac{y_\mu }{\Lambda^2} (\bar L^{(2)}_L\gamma_\mu L_{L})(\bar  e_R\gamma^\mu e^{(1)}_R)
+C^{\mu qqe}_{LR}\frac{y_\mu }{\Lambda^2} (\bar L^{(2)}_L\gamma_\mu Q_{L})(\bar  d_R\gamma^\mu e^{(1)}_R)
 +(\mu\leftrightarrow e)
 \label{op4f0} \\[.2cm]
&+& 
C^{\mu eqq}_{LuQe}\frac{y_\mu }{\Lambda^2} (\bar{L}^{(2)}_L u_R)(\bar{Q}_L  e^{(1)}_R)
+C^{\mu eqq}_{LeQu}\frac{y_\mu }{\Lambda^2} (\bar{L}^{(2)}_L e^{(1)}_R)(\bar{Q}_L  u_R)
 +(\mu\leftrightarrow e)
+\text{h.c.}\,, 
\label{op4f2}
 \eea
 where $y_\mu=\sqrt{2}m_\mu/v$ ($v=246$ GeV) is the SM muon Yukawa coupling and $g$, $g'$ are the SU(2)$_L$ and 
 U(1)$_Y$ couplings
as defined in Appendix \ref{conventions}. We also have $F_L=Q_L,L_L$ and $f_R=u_R,d_R,e_R$ the SM  left-handed SU(2)$_L$ doublets and   
right-handed SU(2)$_L$ singlets respectively.
We only specify  the flavor indices for the $\mu e$ transitions where the superindices $1,2$ corresponds to
$e,\mu$.
We also have factored out a   $y_\mu$ in operators  involving $\bar L^{(2)}_Le^{(1)}_R$ or $L^{(1)}_Le^{(2)}_R$ where
the $\mu e$ transitions have a change of  chirality. Therefore, in  the interchange $\mu\leftrightarrow e$ 
 in ${\cal L}_6$  we keep fixed $y_\mu$.
 
 The four-fermion operators have been separated into two groups: \eq{op4f0} are operators of type $\psi^2\bar\psi^2$
 in Weyl notation, while \eq{op4f2} are operators of type $\psi^4$ (and $\bar\psi^4$).
 They have respectively total helicity $h=0$ and $h=\pm 2$.
This  is important as we will see later for understanding the renormalization mixing.
 As compared to the Warsaw basis~\cite{Grzadkowski:2010es}, we have made the replacement\footnote{Using Fierz identities,  we have the relation ${\cal O}^{(3)}_{lequ} = - 8 {\cal O}_{LuQe}  - 4 {\cal O}_{LeQu}$.}
\be
{\cal O}^{(3)}_{lequ}=(\bar L_L\sigma_{\mu\nu}e_R)  (\bar Q_L \sigma^{\mu\nu}u_R) \ \to\ 
{\cal O}_{LuQe}=(\bar L_L u_R)  (\bar Q_L e_R)\,,\\
\ee
and we have  written differently   the operator (via Fierzing)
\be
{\cal O}_{ledq}=(\bar L_L e_R)(\bar d_R Q_{L})=
-\frac{1}{2}(\bar L_L\gamma_\mu Q_{L})(\bar  d_R\gamma^\mu e_R)
\,,
\ee
to make it clear that this is an operator of the type $\psi^2\bar\psi^2$ with total helicity $h=0$.
 
 It will be important for later  to understand the impact of the  operators \reef{opJJ} to write them
in the unitary gauge:  
 \bea
 -(v+h)^2 &\Bigg[&\frac{C^{\mu e}_{L}+C^{\mu e}_{L3}}{\Lambda^2}\,
\left(\frac{g}{2c_{\theta_W}}
Z^\mu \bar \mu_L\gamma_\mu e_L-\frac{g}{2\sqrt{2}}[W^{+\mu}\bar\nu_{\mu L}\gamma_\mu e_L  + \text{h.c.}] \right)\nonumber\\
&+&\frac{C^{\mu e}_L-C^{\mu e}_{L3}}{\Lambda^2}\,
\left(\frac{g}{2c_{\theta_W}}
Z^\mu \bar \nu_{\mu L}\gamma_\mu \nu_{e L}+\frac{g}{2\sqrt{2}}[W^{+\mu} \bar\nu_{\mu L}\gamma_\mu e_L + \text{h.c.}] \right)\nonumber\\
&+&\frac{C^{\mu e}_R}{\Lambda^2}\, \frac{g}{2c_{\theta_W}} Z^\mu \bar \mu_R\gamma_\mu e_R \Bigg] +(\mu \leftrightarrow e)\,,
\label{unitarygauge}
\eea
which shows that these operators induce a LFV coupling for the $Z$, $W$ and $hW,hZ$.
Custodial symmetry together with a $L\leftrightarrow R$ parity can protect
some of these couplings. For example, as explained in Ref.~\cite{Elias-Miro:2013mua}, these symmetries  can impose $C_{L}+C_{L3}=0$
such that the LVF $Z\mu e$ and $hZ\mu e$ couplings are zero. Therefore, for BSM sectors respecting  these symmetries, 
the only  LFV couplings will involve neutrinos which are difficult to detect, implying 
that no strong bounds can be derived from direct measurements such as $W\to  \mu \bar \nu_e, e \bar \nu_\mu$ or $Z\to  \nu_\mu \bar \nu_e$. 
As we will see, loop effects will be important to get better bounds from  other LFV observables.

\begin{table}[h!]
	\begin{center}
		\begin{spacing}{1.03}
			\begin{tabular}{l l l l l}
				\hline
				& $\mu\rightarrow e\gamma$  & $\mu\rightarrow eee$     & $\mu N \rightarrow eN$                & $h\rightarrow \mu e$ \\ \bottomrule
				$C^{\mu e}_{DB}-C^{\mu e}_{DW}$ & \begin{tabular}[c]{@{}l@{}} 951 TeV\\  (1547 TeV) \end{tabular} & \begin{tabular}[c]{@{}l@{}} 218 TeV\\  (2183 TeV) \end{tabular} & \begin{tabular}[c]{@{}l@{}} 208 TeV\\  (1812 TeV) \end{tabular} &     \\ \hline
				$C^{\mu e}_{DB}+C^{\mu e}_{DW}$ & \colorblock{\begin{tabular}[c]{@{}l@{}} 127 TeV\\  (214 TeV) \end{tabular}} & \colorblock{\begin{tabular}[c]{@{}l@{}} 26 TeV\\  (309 TeV) \end{tabular}} & \colorblock{\begin{tabular}[c]{@{}l@{}} 24 TeV\\  (253 TeV) \end{tabular}} &     \\ \hline
				
				$C^{\mu e}_{R}$        & \colorblockr{\begin{tabular}[c]{@{}l@{}} 35 TeV\\  (59 TeV) \end{tabular}}       & \begin{tabular}[c]{@{}l@{}} 160 TeV\\  (1602 TeV) \end{tabular}   &  \begin{tabular}[c]{@{}l@{}} 225 TeV\\  (1535 TeV) \end{tabular}                   &   \\ \hline
				$C^{\mu e}_{L}+C^{\mu e}_{L3}$       & \colorblockr{\begin{tabular}[c]{@{}l@{}} 4 TeV\\  (7 TeV) \end{tabular}}       & \begin{tabular}[c]{@{}l@{}} 164 TeV\\  (1642 TeV) \end{tabular}   &  \begin{tabular}[c]{@{}l@{}} 225 TeV\\  (1535 TeV) \end{tabular}             &   \\ \hline
				$C^{\mu e}_{L}-C^{\mu e}_{L3}$       & \colorblockr{\begin{tabular}[c]{@{}l@{}} 24 TeV\\  (41 TeV) \end{tabular}}       & \colorblock{\begin{tabular}[c]{@{}l@{}} 35 TeV\\  (421 TeV) \end{tabular}}    &  \colorblock{\begin{tabular}[c]{@{}l@{}} 50 TeV\\  (395 TeV) \end{tabular}} &           \\ \hline
				$C_{LuQe}^{\mu e tt}$      & \colorblock{\begin{tabular}[c]{@{}l@{}} 304 TeV\\  (510 TeV) \end{tabular}}     &  \colorblock{\begin{tabular}[c]{@{}l@{}} 63 TeV\\  (735 TeV) \end{tabular}} &  \colorblock{\begin{tabular}[c]{@{}l@{}} 59 TeV\\  (604 TeV) \end{tabular}}  &      \\ \hline
				$C_{LeQu}^{\mu e tt}$      & \colorblockp{\begin{tabular}[c]{@{}l@{}} 80 TeV\\  (141 TeV) \end{tabular}}     &  \colorblockp{\begin{tabular}[c]{@{}l@{}} 14 TeV\\  (209 TeV) \end{tabular}} &  \colorblockp{\begin{tabular}[c]{@{}l@{}} 5 TeV\\  (57 TeV) \end{tabular}}  &        \\ \hline
				$C_{LL(RR),LR(RL)}^{\mu e e e}$  &    & \begin{tabular}[c]{@{}l@{}} 207,174 TeV\\ (2070,1740 TeV) \end{tabular} &      &              \\ \hline
				$C_{LL,RR,LR}^{\mu e u u}$        &   &   & \begin{tabular}[c]{@{}l@{}} 352 TeV\\  (2693 TeV) \end{tabular}   &                      \\ \hline
				$C_{LL,RR,LR}^{\mu e d d}$        &   &   &  \begin{tabular}[c]{@{}l@{}} 376 TeV\\ (2725 TeV) \end{tabular}   &                         \\ \hline
				$C_{LR}^{\mu d d e}$        &   &   &  \begin{tabular}[c]{@{}l@{}} 18 TeV\\ (164 TeV) \end{tabular}   
				&                        \\ \hline
				$C_{LL,RR,LR,RL}^{\mu e \tau \tau}$        &       &  \colorblock{\begin{tabular}[c]{@{}l@{}} 14,16,14,16 TeV\\  (174,194,174,194 TeV) \end{tabular}}   &  \colorblock{\begin{tabular}[c]{@{}l@{}} 22 TeV\\  (200 TeV) \end{tabular}} 
				&                         \\ \hline
				$C_{LL3}^{\mu e \tau \tau}$        &      &  \colorblock{\begin{tabular}[c]{@{}l@{}} 20 TeV\\  (247 TeV) \end{tabular}}  & \colorblock{\begin{tabular}[c]{@{}l@{}} 55 TeV\\  (476 TeV) \end{tabular}}  
				&                         \\ \hline
				$C_{LL,RR,LR,RL}^{\mu e tt}$      & \colorblockp{\begin{tabular}[c]{@{}l@{}} 122 TeV\\  (214 TeV)\end{tabular}}     &  \colorblockp{\begin{tabular}[c]{@{}l@{}} 21 TeV\\  (317 TeV) \end{tabular}}  &  \colorblock{\begin{tabular}[c]{@{}l@{}} 22,32,32,22 TeV\\  (200,290,290,200 TeV)\end{tabular}} 
				&         \\ \hline
				$C_{LL3}^{\mu e tt}$      & \colorblockp{\begin{tabular}[c]{@{}l@{}} 230 TeV\\  (401 TeV)\end{tabular}}     &  \colorblockp{\begin{tabular}[c]{@{}l@{}} 41 TeV\\  (592 TeV) \end{tabular}}  &  \colorblock{\begin{tabular}[c]{@{}l@{}} 100 TeV\\  (851 TeV) \end{tabular}}  &        \\ \hline
				$C_{LL,RR,LR,RL}^{\mu e bb}$      &      &  \colorblock{\begin{tabular}[c]{@{}l@{}} 14,16,14,16 TeV\\  (174,194,174,194 TeV) \end{tabular}}  &  \colorblock{\begin{tabular}[c]{@{}l@{}} 22 TeV\\  (200 TeV) \end{tabular}}  
				&         \\ \hline
				$C^{\mu e}_y$    & \colorblockr{\begin{tabular}[c]{@{}l@{}}  4 TeV\\  (6 TeV) \end{tabular} }     
				&    \colorblockr{\begin{tabular}[c]{@{}l@{}}  1 TeV\\  (9 TeV) \end{tabular} }   &              \colorblockr{\begin{tabular}[c]{@{}l@{}}  1 TeV\\  (7 TeV) \end{tabular} }            & 0.3 TeV   \\ \hline
			\end{tabular}
			\caption{Present (future) lower 
				bounds on $\Lambda$ of the dimension-six operators  \reef{opJJ}-\reef{op4f2} from
				the different LFV violating processes. We have fixed 
				the Wilson coefficient $C_i=1$,  turning each one by one. We show the bound in black,  \colorblock{blue}, \colorblockp{purple} and \colorblockr{red} depending on whether the Wilson coefficients contribute to the observables at the tree-level, one-loop single-log, two-loop double-log or  two-loop  single-log order, respectively. Blank spaces correspond to contributions that we estimate to be  too small to be competitive with existing bounds.}
			\label{bounds}
			\vspace{-2em}
		\end{spacing}
	\end{center}
\end{table}

\subsection{LFV experimental constraints at the tree level}
\label{soa}

The Wilson coefficients of the operators \reef{opJJ}-\reef{op4f2} are constrained from the 
LFV observables of  Table~\ref{tabbb}.
In this section we want to review the bounds obtained from a tree-level analysis.
 The result is presented in Table~\ref{bounds} in black color
 (see Refs.~\cite{Crivellin:2013hpa,Pruna:2014asa,Pruna:2015jhf,Crivellin:2017rmk} for previous analysis).
   The aim of this analysis is to understand  which   loop effects mixing the different Wilson coefficients
of    \reef{opJJ}-\reef{op4f2}
   are important to consider  in order to obtain better   bounds.
   These loop calculations and the discussion of how much the   bounds are improved will be presented in the next sections.

 The current constraint on the dipoles  $C^{e\mu,\mu e}_{DW,DB}$ of \eq{opD} is dominated by the  
experimental bound on BR$(\mu\rightarrow e \gamma)$ at which  they enter at tree level (see \eq{dip1}).
This leads to
  \be
\frac{1}{\Lambda^2} \sqrt{ |C^{\mu e}_{DW} - C^{\mu e}_{DB}|^2 + |C^{e \mu}_{DW} - C^{e \mu}_{DB}|^2} \lesssim 
1\cdot 10^{-6}\frac{1}{{\rm TeV}^2}\,,
\label{CDWconst}
  \ee
and  will improve by almost an order of magnitude  in the future.
 \eq{CDWconst} provides a very strong constraint that shows that renormalization effects to $(C_{DW} - C_{DB})$
arising  from other Wilson coefficients $C_i$ could also place  strong constraints on these latter.  
These effects can be  important  even when they arise at the two-loop level, 
e.g. $\Delta (C_{DW} - C_{DB})\sim {C_i}/{(16\pi^2)^2}\sim 10^{-4}$,
where we estimate to obtain $C_i/\Lambda^2\lesssim 1/(6\ {\rm TeV})^2$.
Therefore an analysis of anomalous dimension mixings into $C_{DW,DB}$ is needed up to the two-loop level.
 The  Wilson coefficient $(C_{DW} - C_{DB})$ also enters at tree level in the observables
   $\mu\rightarrow 3e$ and $\mu N \rightarrow e N$ (see \eq{BRmu333} and \eq{BRmuNeN} respectively), 
   but the present constraints are not so competitive, as shown in Table~\ref{bounds}.
Nevertheless,  the spectacular prospects of improvement by four orders of magnitude in each of them  will lead to an improved bound on $(C_{DW} - C_{DB})$   that could overcome that  from  $\mu\rightarrow e \gamma$.
 
 Let us consider now the four-fermion interactions \reef{op4f0}  and \reef{op4f2}.
 In particular, the operators of type  $\mu \bar e\bar ee$ enter at tree level to $\mu\to 3 e$ and provide the limits
\be
\frac{C^{\mu eee}_{LL,RR}}{\Lambda^2} \lesssim 2.3\cdot 10^{-5}\frac{1}{{\rm TeV}^2}\,, \qquad \frac{C^{\mu eee}_{LR,RL}}{\Lambda^2} \lesssim 3.3\cdot 10^{-5}\frac{1}{{\rm TeV}^2}\,. 
\label{C4fconst}
\ee
They are clearly not as strong as that in \reef{CDWconst}, but  in the near future this bound
 is expected to  improve by two orders of magnitude, being then comparable to \eq{CDWconst}.
Therefore loop effects mixing other Wilson coefficients into $C^{\mu eee}_{LL,RR,LR,RL}$ 
 should also be considered.
 As we will see, however,  all the relevant LFV Wilson coefficients enter into the anomalous dimensions
 of $C^{\mu eee}_{LL,RR,LR,RL}$ at the one-loop level, making two-loop effects not necessary.
Similar conclusions can be reached for four-fermion interactions of type $\mu \bar e\bar uu$ and $\mu \bar e\bar dd$ 
that are constrained by $\mu N\to eN$ at a similar level as in \eq{C4fconst}.

The  Wilson coefficients of \eq{opJJ}  give rise to the LFV  $Z\mu e$ coupling  (see \eq{unitarygauge}),
and therefore can enter at tree level into the observables $\mu\to 3e$ and $\mu N\to eN$ (see \eq{BRmu333} and \eq{BRmuNeN} respectively),
providing a bound similar to \reef{C4fconst}.
There are also direct bounds from  $Z\to\mu e$ 
 but these are not so competitive (one gets constraints of order $\Lambda\gtrsim 5$ TeV).
 It is important to remark again  that the induced $Z\mu e$ coupling is proportional 
to the combination $(C^{\mu e}_{L}+C^{\mu e}_{L3})$, and therefore tree-level constraints
from $\mu\to 3e$ and $\mu N\to eN$ are only on this specific combination. 
As we already discussed,  the  orthogonal combination $(C^{\mu e}_{L}-C^{\mu e}_{L3})$ only
enters at tree level into LFV gauge boson couplings involving neutrinos, implying that no relevant bounds
on this combination can be placed at this order.
Loop effects however can be important as we will discuss in the next section.
The parametrization of the bounds using  these two particular combinations has also a theoretical motivation, since,
 as we already mentioned,  BSM models can give sizable contributions to
$(C^{\mu e}_{L}-C^{\mu e}_{L3})$ but not to $(C^{\mu e}_{L}+C^{\mu e}_{L3})$
as this latter is protected by a custodial symmetry~\cite{Elias-Miro:2013mua}. We provide an example in Section~\ref{singlet}.

The last operator to consider is \reef{opy} that can only enter at tree level in Higgs physics, specifically in $h\to \mu e$.\footnote{These Wilson coefficients enter as 
BR$\left(h\rightarrow \mu e\right)=m_H m^2_\mu v^2 [\abs{C_{y}^{\mu e}}^2+\abs{C_{y}^{e\mu}}^2 ]
/(8\pi\Gamma_H\Lambda^4)$.
}
The bound (see Table~\ref{bounds}) is very weak, therefore it is expected that 
bounds from other observables, where $C^{\mu e,e\mu}_{y}$ can enter via mixing at the loop level, can be potentially stronger. We will indeed see that this operator enters at the two-loop level into $\mu\to e\gamma$ and gets a much more severe constraint.

We conclude that there are  some Wilson coefficients whose  loop effects can be potentially relevant for the observables  
$\mu\to e\gamma$, $\mu\to 3e$
and $\mu N\to eN$. 
Below we will calculate the anomalous dimensions at the two-loop level for these $C_i$ that will allow us to obtain
new competitive bounds.

\section{$\mu\to e\gamma$ in the SM EFT}
 \label{muegz}

The $\mu\to e\gamma$ process  arises from the Lagrangian term
\be
 -\frac{4 G_F}{\sqrt{2}}m_\mu\left[  d_{\mu e}\, \bar{\mu}_L \sigma^{\mu\nu} e_R F_{\mu\nu}\,+\, 
d_{e \mu}\, \bar{e}_L \sigma^{\mu\nu} \mu_R F_{\mu\nu}
\,+\, \text{h.c.}\,\right],
\label{eq:FVdipole}
\ee
where $d_{\mu e,e \mu}$ must be evaluated at the muon mass scale.
The branching ratio is given by \cite{Kuno:1999jp}
\be
\text{BR}\,(\mu\to e\gamma) \,=\, 384\pi^2\,\left(|d_{\mu e}|^2+|d_{e\mu}|^2\right)\,, \label{brmue}
\ee
where the large factor is because the LFV decay is a two-body process, while the dominating muon decay channel is a three-body one.  This decay process is tightly constrained and further sensitivity is expected in the near future, see Table \ref{tabbb}.

At tree level the only Wilson coefficients of the SM EFT  that enter into $d_{\mu e}$ are the dipoles $C^{\mu e}_{DW,DB}$:
\be
d_{\mu e} = \frac{e\, v^2}{2\Lambda^2}\left(C_{DW}^{\mu e} - C_{DB}^{\mu e}  \right)\,,  \label{dip1}
\ee
where $e=g \sin \theta_W$,
and similarly for $d_{e\mu }$ by exchanging $\mu\leftrightarrow e$.

We are interested in operator mixing into the dipoles \reef{eq:FVdipole}. 
There are various effects to be considered: $(i)$  finite matching  contributions from a  new physics scale $\Lambda$,
$(ii)$ RG mixing from $\Lambda$ to the electroweak scale $m_W$, $(iii)$ finite  threshold corrections arising from  integrating out the $W$, $Z$, Higgs and the heavy SM fermions, and $(iv)$ 
RG mixing from  the electroweak scale $m_W$ to $m_\mu$.
At the leading order (one-loop level) the UV matching threshold $(i)$ and SM IR matching $(iii)$  lead to finite contributions to the dipoles. 
These  type of  corrections  are model dependent, and can (partially) cancel against each other
as they  involve  rational coefficients, apart from some overall couplings.
 We will show examples in which this occurs in Section~\ref{singlet}. 
Here we are instead interested in contributions from RG running, which correct the tree-level dipoles by logarithms $\ln(\Lambda/m_W)$, raised to a suitable power,  which are hardly possible to cancel against matching contributions of either type $(i)$ and $(iii)$. 
In particular we are interested in the leading RG mixings from any operator, other than the dipoles, into the dipoles. 
Our aim is to understand to which physics a precise measurement of \reef{brmue} is sensitive. Therefore, 
we will only keep track of the leading RG mixing contribution from any dimension-six operator into \reef{eq:FVdipole}.

At the one-loop level, selection rules, mainly based on spurious SUSY 
\cite{Elias-Miro:2014eia} or helicity arguments \cite{Cheung:2015aba}, 
dictate the anomalous dimension  mixing terms ${\cal O}_i\to {\cal O}_j$.
Of particular importance is  the selection rule $\Delta n\geq |\Delta h|$ 
\cite{Cheung:2015aba}, where $\Delta n=n_f-n_i$ and $\Delta h=h_f-h_i$, being $n$ ($h$) the number of fields (helicity) of the operator.  There is only one exception to this rule, $\bar \psi^2\psi^2\leftrightarrow \psi^4$ only when the loop involves the Yukawa product $y_uy_d$ or $y_uy_e$.
In the particular case of dipole operators that have $n=4,|h|=2$,
   the only operators that can mix  into them  are those with
 helicity $|h|=2$. This reduces to the   four-fermion operators  in \reef{op4f2}, apart from the orthogonal combination 
$(C_{DW}^{\mu e} + C_{DB}^{\mu e})$.
The one-loop RG mixing from ${\cal O}^{\mu e qq}_{LeQu}$  however is trivially absent because $\bar L^{(2)} e^{(1)}$ is external to the loop calculation   and does not have the  dipole structure (and similarly for  ${\cal O}^{ e \mu  qq}_{LeQu}$).   
This argument leaves out only the following possibility for one-loop leading-log contributions:
\be
C_{LuQe}\  \longrightarrow \  C_{DW,DB} \, ,    \label{leadingone}
\ee
where we suppressed flavor indices.
The calculation of the one-loop RG mixing in \reef{leadingone}  was done in \cite{Jenkins:2013wua} and  recently revisited in \cite{Baratella:2020lzz} with on-shell methods. The result is presented in Section~\ref{oneloop}.

At the two-loop level, there are other various dimension-six operators that can mix with the dipoles. 
We can classify them according to  a logarithmic  expansion.
At leading order we have  two-loop double-log contributions proportional to
\be
\frac{C_i C_j}{ (16\pi^2)^2 }  \ln^2(\Lambda/m_W)\,, \quad i\neq j\,. 
\label{typetwo}
\ee 
These contributions arise from mixings of the type $\O_i  \stackrel{\text{1-loop}}{ \longrightarrow} \O_j   \stackrel{\text{1-loop}}{ \longrightarrow}   \O_{DW,DB}$.
Again, selection rules \cite{Elias-Miro:2014eia,Cheung:2015aba}
only allow  operator mixing between those with equal $h$, in particular,
${\cal O}_{LeQu} \rightarrow {\cal O}_{LuQe} \rightarrow {\cal O}_{DW,DB}$, with the exception of 
$\bar \psi^2 \psi^2 \rightarrow \psi^4 \rightarrow \psi^2 \phi F$ where the first step can occur if it involves the product of Yukawa couplings $y_uy_d$ or $y_uy_e$.  
There are double logs of the type  $C_i^2 \ln^2(\Lambda/m_W) /(16\pi^2)^2$ as well.  However these are sub-leading corrections to  already existing one-loop effects.
The second type of corrections are two-loop single-log contributions    
\be
\frac{C_i}{(16\pi^2)^2} \ln(\Lambda/m_W)\,.  
\label{typeone}
\ee
There are only  three of  such two-loop mixings: $\psi^2 \phi^3,\psi^2\bar \psi^2\rightarrow \psi \phi F$, which were  computed in \cite{Panico:2018hal,EliasMiro:2020tdv}, and 
$\bar \psi \gamma \psi H^\dagger D H \longrightarrow F_{\mu\nu}\bar \psi H \psi$, which will be calculated here for the first time. 

\begin{figure}
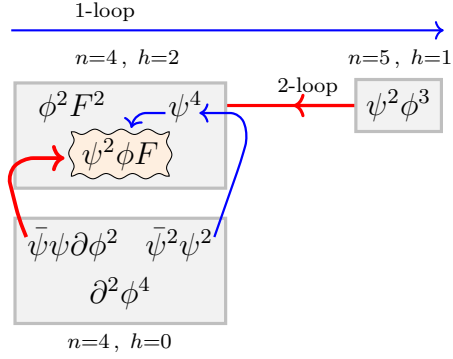

	\centering
 \begin{minipage}[h][4.65cm][t]{0.25\linewidth} \vspace{-.12cm}
 \dMaster 
   \end{minipage}  
	\caption{ Structure of the  RG mixings from dimension-six operators into LFV dipoles. 
One-loop RG mixing can only proceed from left to right, generating mixing between operators belonging to different helicity boxes (shown in grey), and within operators belonging to the same helicity box. One-loop mixing from the category   $(n,h)=(4,0)$ into $(4,2)$ is forbidden by spurious SUSY~\cite{Elias-Miro:2014eia} or helicity~\cite{Cheung:2015aba} selection rules,  except for the  non-holomorphic contribution proportional to $y_uy_d$ or $y_uy_e$, 
	which mixes the $\psi^4$ and $\bar\psi^2\psi^2$ four-fermion operators at one loop.
	In {\blue blue} and  {\red red} we indicate the relevant RG mixings for the LFV dipoles at one- and two-loop precision, respectively. 
 }
 	\label{figdiag}
\end{figure}

We summarize the structure of the one- and two-loop mixing pattern into dipoles in Figure \ref{figdiag}.
 In the next section we present the calculation of the relevant anomalous dimensions  of dipole Wilson coefficients up to the
  two-loop level. 
  Our  task is greatly  simplified by using on-shell methods as we explain next.
 
\section{The on-shell way}
\label{twoloopd}

Let us start presenting the basic concepts necessary for the  computation of anomalous dimensions from on-shell amplitudes. 
Our calculations are  based on the  
formula~\cite{Caron-Huot:2016cwu}
\be
\bra{\vec{n}}{\cal O}_j \ket{0}^{(0)} \gamma_{ji}^{(1)} =- \frac{1}{\pi}\,  \bra{\vec{n}}{\cal M}\otimes {\cal O}_i \ket{0}^{(0)}  \, , \label{chw}
\ee
which allows to compute the leading correction to the anomalous dimension matrix 
 $\gamma_{ji}^{(1)}$ with $i\neq j$.
On the left-hand side (l.h.s.) of \reef{chw} we have a  form factor  computed  in the free theory, i.e. at zero order in the coupling, denoted as minimal form factor. 
For instance, the minimal form factors of the  dipoles are given by
\bea
\bra{   1_{L_l}^-     2_B^-  3_e^-  4_{H_k}} {\cal O}_{DB}\ket{0}^{(0)} & =&  2\sqrt{2}g^\prime  y_\mu \langle 1 2 \rangle\langle 2 3  \rangle \eps_{lk}  \label{fdb} \, ,  \\
\bra{  1_{L_l}^-   2_{W^\alpha}^-   3_e^-  4_{H_k}   } {\cal O}_{DW}\ket{0}^{(0)} &=&  2\sqrt{2}  g  y_\mu   \langle 1 2 \rangle\langle 2 3  \rangle \eps_{lk'} (\tau^\alpha)^{k'}_{k} \label{fdw}\, ,
\eea
where $\eps$ and $\tau$ are  tensors of SU(2)$_L$.\footnote{Further details on notation 
are given in Appendix~\ref{conventions}.}
The  form factors $\bra{\vec{n}}{\cal O}_j \ket{0}^{(0)}$ are polynomials of the kinematical variables $\{\ket{i},|i]\}$, i.e. the spinor-helicity variables.\footnote{Form factors are defined by $F_{\cal O}(1\cdots n;q)= \int d^dx \langle 1 \cdots n | {\cal O}(x) | 0 \rangle e^{i q x}$, where ${\cal O}(x)$ is a local operator and $\langle 1 \cdots n | $ is an ``out" (outgoing) state. In the examples considered in this paper  we can smoothly send $q\rightarrow 0$. In this limit the form factor is equivalent to a scattering amplitude with an insertion of the  EFT operator $ \int d^dx {\cal O}(x)$ and with all particles outgoing. }
On the right-hand side (r.h.s.),  the symbol `$\otimes$' denotes a $d\text{LIPS}$ phase-space integration over the intermediate states $\sum_{\vec{m}}\ket{\vec{m}}\bra{\vec{m}}$ connecting the on-shell  scattering amplitude  ${\cal M}(\vec{n}\leftarrow \vec{m})$ and the on-shell form factor $\bra{\vec{m}}{\cal O}_i\ket{0}$.
The integral over the phase space is often referred to as \emph{unitarity cut}, or \emph{cut} for short. When evaluating the r.h.s. with Feynman diagrams, a cut corresponds to putting on shell the  internal propagators crossing the cut, which has the net effect of cutting the Feynman diagram into lower-point diagrams sewed together through the phase-space integral. 
The upper script $(0)$ means that the r.h.s. of \reef{chw} is computed  to leading non-trivial order so that the unitarity cut  $\bra{\vec{n}}{\cal M}\otimes {\cal O}_i \ket{0}^{(0)}$ leads to a polynomial in the kinematical variables, matching the left-hand side of \reef{chw}. Further details and examples can be found in \cite{Caron-Huot:2016cwu,EliasMiro:2020tdv,Bern:2020ikv}.

\subsection{One-loop mixing into LFV dipoles}
\label{oneloop}

When only two particles cross the cut in the r.h.s. of \reef{chw}, and when only four particles are involved in either the form factor or the amplitude, 
we are left with the following phase-space integral~\cite{Caron-Huot:2016cwu}
\be
\begin{minipage}[h][1.5cm][t]{0.19\linewidth} \vspace{.25cm} \oneloopUn \end{minipage}  =  - \frac{1}{8\pi^2} \int d\Omega_2 \, M(12 ;  x y ) \, F_{{\cal O}_i}( x y  34)  \, ,  \label{one}
\ee
where the amplitude  $M$ describes the $x+y\rightarrow 1+2$ scattering process, and as usual we have extracted a total delta function $\langle 12|{\cal M}|xy\rangle= (2\pi)^4\delta^{(4)}(p_1+p_2-p_x-p_y)M(12;xy)$.
The particles of the form factor  $F_{{\cal O}_i}(1 2 \cdots n)\equiv \langle 1 2 \cdots n|{\cal O}_i | 0 \rangle$ are all outgoing.
The integral can be easily performed if we  write the spinors in the integrand  in a basis spanned by two of the external spinors
\be
\begin{pmatrix}
	 |x\rangle  \\
	 |y\rangle
\end{pmatrix} = 
\begin{pmatrix}
	\cos {\th} & -e^{i\phi}\sin {\th} \\
	e^{-i \phi } \sin \th  &  \cos \th
\end{pmatrix}
\begin{pmatrix}
	 |1\rangle \\
	|2\rangle  
\end{pmatrix} \label{rotone} \, , 
\ee
with the measure  given by $d\Omega_2\equiv \frac{d\phi}{2\pi}  2\cos\th \sin\th d \th $.
The rotation  of  $|x],|y]$ into $|1],|2]$ is obtained by complex-conjugation of \reef{rotone}. When identical particles cross the cut in \reef{one}, one should include an extra combinatorial symmetry factor of 1/2! in the phase space.

As we already explained in the previous section, at the one-loop level we only have \eq{leadingone}.
This   computation was done  in \cite{Baratella:2020lzz} using unitarity cuts,  agreeing with \cite{Jenkins:2013wua,Panico:2018hal}. As a warm-up,  we review next such computation. 
For the case at hand,   \eq{one} is given by
\be
\begin{minipage}[h][2.6cm][t]{0.28\linewidth}  \OneLoopFirst \end{minipage}  
\ee
where on the l.h.s. the gauge boson must be attached in all possible ways to the Higgs (dashed line) or the fermion lines (solid).
This results in a  two-to-two amplitude given by
\be
M(2 4_k ; x_a y) = - \sqrt{2} y_t \left( Y_{t_R} \frac{[xy]^2}{[x2][y2]} + Y_H \frac{[xy][4x]}{[42][x2]}  \right)  {\cal T}^a_k\,, \label{Mone}
\ee
where ${\cal T}^a_k = g^\prime \delta^a_k$ is the SU(2)$_L$ tensor arising from the contraction of left-handed doublets. On the right-hand side we have the form factor given by
\be
F_{{\cal O }_{LuQe}^{e\mu  t t }}(x_a y 1_l 3)
=  - y_\mu\expec{1y}\expec{x3} \eps_{la} \, .   \label{Fone} 
\ee
It is now a straightforward matter to plug \reef{Mone} and \reef{Fone} into \reef{one}, perform the spinor rotations \reef{rotone} and a few elementary integrals, leading to 
\be
\underbrace{\phantom{\Big(}2\sqrt{2} g^\prime y_\mu \expec{12}  \expec{23} \eps_{lk} }_{ \text{dipole}}
 \,  (- y_t) \frac{N_c /2}{(16\pi^2)} \left(  Y_H - 2 Y_{t_R} \right)\,,  \label{res1loop}
\ee
where $N_c=3$ is the number of colors and $y_t$ the top Yukawa coupling. 
We recognize the minimal form factor of the dipole \reef{fdb} and therefore the anomalous dimension is $\gamma_{DB} = \frac{y_t}{2}\frac{N_c}{16\pi^2}(Y_{Q_L} + Y_{t_R})$.

In the case of mixing into $\O_{DW}$, we set hypercharges $Y_{t_R} = 0$ and $Y_H = 1$ and change the SU(2)$_L$ tensor to be ${\cal T}^a_k = g (\tau^\alpha/2)^a_k$ on the amplitude side. Then we get the following result:
\be
\underbrace{2\sqrt{2}  g  y_\mu   \langle 1 2 \rangle\langle 2 3  \rangle \eps_{lk'} (\tau^\alpha)^{k'}_{k} }_\text{dipole} \,  (- y_t)\frac{N_c/4}{(16\pi^2)} .
\ee
From the last expression we recognize the dipole \reef{fdw} and the corresponding anomalous dimension.

\subsection{Two-loop mixing into dipoles}

We want to calculate here the two-loop mixing $\bar \psi \gamma \psi H^\dagger D H \longrightarrow F_{\mu\nu}\bar \psi H \psi$, which is the only one relevant for $\mu\to e\gamma$ not yet calculated.
The two-loop leading-log contributions to the r.h.s. can in principle   involve three-particle cuts or two-particle cuts:
\be
\begin{minipage}[h][1.5cm][t]{0.19\linewidth} \vspace{.25cm} \twoloopgenUn \end{minipage}  \ +  \  \  
\begin{minipage}[h][1.5cm][t]{0.19\linewidth} \vspace{.25cm}  \twoloopgenTres \end{minipage}  \ +  \  \ 
\begin{minipage}[h][1.5cm][t]{0.19\linewidth} \vspace{.25cm}  \twoloopgenDos \end{minipage}  \, .  \label{gentwoloop}
\ee
The first diagram  involves a tree-level amplitude and a tree-level form factor, so that the three-particle cut accounts for the two-loop factor. The second/third diagram involves a tree-level/one-loop amplitude and a one-loop/tree-level form factor which, together with the two-particle cut, make it to  two-loop order. 
Bellow we will show that the  second and third diagrams do not contribute to the $\bar \psi \gamma \psi H^\dagger D H \longrightarrow F_{\mu\nu}\bar \psi H \psi$ mixings because of simple helicity selection rules. 
Thus, all our non-trivial calculations will only involve three-particle cuts. 
For the transition $\bar \psi \gamma \psi H^\dagger D H \rightarrow F_{\mu\nu}\bar \psi H \psi$, in   \reef{gentwoloop} we only need to consider two external particles to the scattering amplitude and form factor. 

The phase-space integral involving the three-particle cuts can be nicely simplified into the following angular integration~\cite{Caron-Huot:2016cwu}
 \be
\begin{minipage}[h][1.5cm][t]{0.17\linewidth} \vspace{.25cm} \twoloopgenUn \end{minipage} = \frac{\langle 12\rangle[12]}{(16\pi^2)^2 }\int d\Omega_3 \, M(12 ;  x y z) \, F_{\O_i}( x y z 34)  \, ,  \label{rot0}
\ee
where the amplitude describes the $x+y+z\rightarrow 1+2$ scattering process at tree level. 
The spinors  in the integrand  can be rotated  in terms of a basis spanned by the two external spinors: 
\be
\begin{pmatrix}
	|x\rangle \\
	|y\rangle \\
	|z\rangle
\end{pmatrix} = 
\begin{pmatrix}
	\cos {\th_1} & -e^{i\phi}\cos {\th_3}\sin {\th_1} \\
	\cos {\th_2}\sin {\th_1} & e^{i\phi}\left(\cos {\th_1}\cos {\th_2}\cos {\th_3}-e^{i\delta}\sin {\th_2}\sin {\th_3}\right) \\
	\sin {\th_1}\sin {\th_2} & e^{i\phi}\left(\cos {\th_1}\cos {\th_3}\sin {\th_2}+e^{i\delta}\cos {\th_2}\sin {\th_3}\right)
\end{pmatrix}
\begin{pmatrix}
	|1\rangle\\
	|2\rangle
\end{pmatrix} \label{rot1} \, , 
\ee
and the  measure   is $
 d\Omega_3 = 4\cos\th_1\sin^3\theta_1d\th_1 2\cos\th_2\sin\th_2d\th_22\cos\th_3\sin\th_3 d\th_3\,\frac{d\delta}{2\pi}\,\frac{d\phi}{2\pi}$.
 In the case that $n$ identical particles cross the cut, one should account for an extra symmetry factor of $1/n!$ in the phase-space integral, as we will see in some examples in the next sections. Note also that \reef{rot0} includes the $-1/\pi$ factor in the r.h.s. of \reef{chw}.

\subsubsection{Top Yukawa $ y_\text{t}^2$ contributions}

We expect this type of contributions to be the dominant ones because they are proportional to   $ N_c y_\text{t}^2$.
 We first explain in detail  the mixing of $\O_{L}^{e\mu}$ into $\O_{DB}^{e\mu}$ through a top loop. 
The three-particle cut is given by 
\be
\begin{minipage}[h][2.5cm][t]{0.31\linewidth}  \TwoLoop \end{minipage}   \, . 
\label{diagtop1}
\ee
The disconnected gauge boson notation means that  the gauge boson must be attached anywhere in the left-hand side of the cut, i.e.
\be
\begin{minipage}[h][2.5cm][t]{0.15\linewidth}   \vspace{.22cm}  \TwoLoopEXone \end{minipage}  = \ \
\begin{minipage}[h][2.5cm][t]{0.12\linewidth}  \vspace{.35cm}   \TwoLoopEXtwo \end{minipage}  +
\begin{minipage}[h][2.5cm][t]{0.15\linewidth}   \vspace{.35cm}   \TwoLoopEXthree \end{minipage}  +
\begin{minipage}[h][2.5cm][t]{0.12\linewidth} \vspace{.36cm} \TwoLoopEXfour \end{minipage}  +
\begin{minipage}[h][2.5cm][t]{0.12\linewidth}  \TwoLoopEXfive \end{minipage}  +
\begin{minipage}[h][2.5cm][t]{0.12\linewidth}  \TwoLoopEXsix \end{minipage}  
 \, . 
\label{discgauge}
\ee
Summing over all such possible attachments of the gauge boson leads to the following tree-level scattering amplitude
\be
M_1(32;  x_ay_bz)=\sqrt{2} y_t y_\mu \left( Y_{\mu_R} \frac{\expec{yz}}{[x2][32]} -  Y_{t_R}\frac{\expec{x3}}{[y2][z2]}- Y_H \frac{\expec{z3}}{[y2][x2]} \right) {\cal A}_{ba}   \, , \label{amp1}
\ee
where ${\cal A}_{ba}=g^{\prime}\eps_{ba}$ is a SU(2)$_L$ tensor arising from the contraction of the left-handed doublets.
We have computed this amplitude using BCFW recursion relation~\cite{Britto:2004ap,Britto:2005fq} with a Risager shift~\cite{Risager:2005vk}.\footnote{
We emphasise that in \reef{amp1}  the particles  are not outgoing, instead \reef{amp1} describes the actual physical process $ x_ay_bz\rightarrow 32$. For crossing fermions we used the same rules as in~\cite{Caron-Huot:2016cwu}, which can be derived by gluing factorized amplitudes that exchange a fermion, see also appendix of~\cite{Baratella:2020lzz}. }
In \reef{amp1} we have eliminated the hypercharge of the left handed fermions in favour of the Higgs hypercharge and right handed fermions hypercharge. 
The hypercharges  are given by $(Y_{\mu_R},Y_{t_R},Y_H)=(-1,2/3,1/2)$.\footnote{The hypercharge of the left-handed fermions has been expressed in terms of those of the right-handed fermions and the Higgs $Y_{\psi_R}+Y_{\overline{\psi}_L}+Y_H=0$.} 
The tree-level form factor on the r.h.s. of the cut  is given by 
\be
F_{\O_{L}^{e\mu}}(x_ay_bz1_l4_k)= -2y_t \frac{\expec{14}[4x]}{\expec{yz}} \, {\cal B}_{kl}^{ba}  \,, \label{ff1}
\ee
where the SU(2) tensor is  $ {\cal B}_{kl}^{ba}= - \delta_{k}^b \delta_{l}^a$. 
Next we plug  \reef{amp1} and \reef{ff1} into \reef{rot0}, perform the rotations \reef{rot1}\footnote{Note that for the choice of  momenta assignments in \reef{diagtop1} the $\expec{12}[12]$ factor in \reef{rot0} should now be $\expec{32}[32]$.}, and after some simple algebra and elementary integrations we are led to 
\be
\underbrace{\phantom{\Big(}2\sqrt{2} y_\mu g^\prime \expec{12}  \expec{23} \eps_{lk} }_{ \text{dipole}}
 \,  \frac{ -N_c  y_t^2}{(16\pi^2)^2} \left(   Y_{\mu_R}+ 2 Y_H \right)  \, .  \label{res1}
\ee
We recognize the minimal form factor of the dipole  \reef{fdb}.

Regarding the two-particle cuts,  we will next argue that  the second and third diagrams of \reef{gentwoloop} vanish. 
For the sake of this argument, we will now consider all  the particles of  the amplitude in the r.h.s. of the cut as outgoing. 
Eq.~\reef{one} involves a physical $\{in\}\rightarrow\{out\}$ amplitude, 
but as customary in the scattering amplitudes literature,
we will cross all particles of the amplitude to be outgoing. 
Consider first the second diagram of \reef{gentwoloop}:  the only potential contribution to such cut involves a tree-level amplitude with all negative helicity (outgoing) particles, which vanishes in the SM because such maximal helicity violating amplitude does not exist.  
Next we consider the third diagram in \reef{gentwoloop}, which for our case is
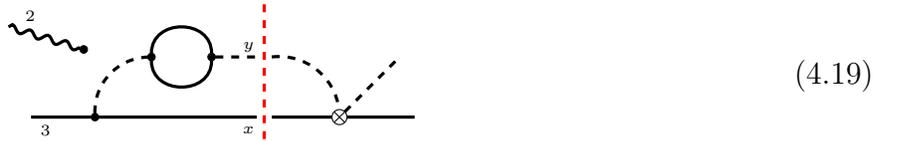
\be
\begin{minipage}[h][2.cm][t]{0.31\linewidth}  \OneLoop \end{minipage}  
\ee
To obtain the l.h.s. scattering amplitude, one must sum over all the diagrams with the gauge boson  attached anywhere on the particles to the left of the cut. 
If the gauge boson is attached to the lepton fermion line, the internal Higgs line goes on shell and the diagram factorizes into  a one-loop dressing of the Higgs line and an all negative helicity tree-level diagram that vanishes. If the gauge boson is attached elsewhere (i.e. Higgs lines or top loop), the sum of the diagrams is equal  to $\expec{3x}f(2,y)$, with $f$ a function that depends only on the 2 and $y$ spinors.  This amplitude vanishes because it is not possible to build an invariant with $\expec{2y}$ and $[2y]$ which has zero helicity for the scalar (of momenta $p_y$) and helicity $h=-1$ for the gauge boson (of momenta $p_2$). 

Therefore the contribution of $\O_{L}^{e\mu}$ to  $\O_{DB}^{e\mu}$ proportional to $N_c y_t^2$ comes only from the three-particle cut \reef{diagtop1} and is given by
 $
- N_c y_t^2 \left(  Y_{\mu_R}+ 2 Y_H \right)/(16\pi^2)^2
$.
This particular contribution is actually zero for the SM hypercharges $Y_{\mu_R}+ 2 Y_H=0$. This is a surprising accidental zero in the anomalous dimension matrix of the SM EFT operators. However, below we will show that other mixings of the type $\bar\psi \gamma \psi HDH\, \substack{\text{2-loop} \\ \longrightarrow}  \, \text{dipoles}$ are not zero.

The  computation of the mixings from  ${\cal O}^{e\mu}_{R}$ to the dipoles is quite analogous: only three-particle cuts matter (for exactly the same reasons as before). The only cut is given by
\be
\begin{minipage}[h][2.5cm][t]{0.31\linewidth}  \TwoLoopR \end{minipage}   \label{diag22}
\ee
Again, one must sum over all the diagrams where the gauge boson is attached anywhere on the l.h.s., so that we are led to the following five-point amplitude
\be
M_2(1_l 2;  x y_bz)=\sqrt{2} y_t y_e \left( - Y_{e_R} \frac{\expec{yz}}{[12][x2]} + Y_{t_R}\frac{\expec{1x}}{[y2][z2]} - Y_H \frac{\expec{zx}}{[12][y2]} \right) {\cal A}_{bl}   \, , \label{amp2}
\ee
while the form factor is 
\be
F_{\O_{R}^{e\mu}}(xy_b z3 4_k)= -2y_t \frac{\expec{34}[4x]}{\expec{yz}} \, {\cal B}_{k}^{b}  \, .  \label{ff2}
\ee
It is again straightforward to perform the rotation \reef{rot1} and integrations in \reef{rot0}.\footnote{In this case, given the momenta assignments in \reef{diag22}, we have to consider a factor $\expec{12}[12]$.} The SU(2) tensors ${\cal A}$ and ${\cal B}$ for the $C^{e\mu}_R$ contributions are given in the table of \reef{tensors} below. 
We get 
\be
\underbrace{\phantom{\Big(}2\sqrt{2} y_\mu g^\prime \expec{12}  \expec{23} \eps_{lk} }_{ \text{dipole}} \ (Y_{e_R}-Y_H) y_t^2  y_e/y_\mu  \, . 
\ee

It is now  straightforward  to generalize our computation for the remaining mixings from the operators $\O_{L}$, $\O_{L3}$ and $\O_{R}$ to the   dipoles. 
There are a few minimal changes that we now specify. First, we need to determine the tensors ${\cal A}$ and ${\cal B}$. We find:
{ \renewcommand{\arraystretch}{1.4} \renewcommand\tabcolsep{6.pt}
 \be
\begin{array}{L{1.8cm} | L{4cm}  L{4.5cm} L{3.5cm} }  
  ${\cal A} \times {\cal B}$ & ${\cal O}_{L} $ & ${\cal O}_{L3}$  &  ${\cal O}_R$  \\  \hline
 ${\cal O}_{DB} $  & $g^\prime \eps_{ba} \times (- \delta_{k}^b \delta^{a}_l) $ & $g^\prime \eps_{ba} \times (\tau^\beta)_{k}^b (\tau^\beta)^{a}_l $ & $g^\prime  \eps_{bl} \times \delta_k^b  $\\ 
 ${\cal O}_{DW} $  & $g \eps_{ba'} (\tau^\alpha/2)^{a'}_a \times (- \delta_{k}^b \delta^{a}_l) $ & $g \eps_{ba'} (\tau^\alpha/2)^{a'}_a \times (\tau^\beta)_{k}^b (\tau^\beta)^{a}_l $   & $ g \eps_{bl'} (\tau^\alpha/2)^{l'}_l \times \delta_k^b $   \\  
 \end{array} \label{tensors}
\ee
}

\noindent
We remark that the ${\cal A}$ and ${\cal B}$ tensors for $\O_L$ and $\O_{L3}$ should be used  in \reef{amp1} and \reef{ff1}, while those of $\O_R$ are used in 
\reef{amp2} and \reef{ff2}. 
Next, we note  that for the $\O_{DW}$ computation, apart from using the correct ${\cal A}$ tensors given in  \reef{tensors}, we also set the hypercharges  $Y_{e_R}=Y_{\mu_R}=Y_{t_R}=0$, $Y_H=1$ in the amplitudes (\ref{amp1},\ref{amp2}), and the SU(2) generators are already included in ${\cal A}$.

Finally, putting all the contributions together we get the following anomalous dimension matrix 
$
( \gamma_{C_{DB}^{e\mu}} , \,     \gamma_{C_{DW}^{e\mu}}  )^T=
 \gamma_{D1}\cdot 
(C_{L}^{e\mu} ,\,   C_{L3}^{e\mu} , \,    C_{R}^{e\mu})^T
$ with 
\be
\gamma_{D1}= \frac{N_c y_t^2}{(16 \pi^2)^2} 
\left( \begin{array}{ccc}
	 0 & 0 & -3 y_e/(2y_\mu)  \\[.2cm]
	1 & - 1 & y_e/(2y_\mu) 
\end{array}\right)
\approx \frac{N_c y_t^2}{(16 \pi^2)^2} 
\left( \begin{array}{ccc}
	0 &  0 &0 \\[.2cm]
	1 & -1  & 0 
\end{array}\right)  \, ,  \label{use1}
\ee
where in the last step we have neglected the terms of order $O(y_e/y_\mu) \approx 0$. The contributions from $C_L^{\mu e}$, $C_{L3}^{\mu e}$ and  $C_R^{\mu e}$ into the dipoles  can be easily obtained by exchanging $\mu \leftrightarrow e$ in the amplitudes and form factors. We find
$
( \gamma_{C_{DB}^{\mu e}} , \,     \gamma_{C_{DW}^{\mu e}}  )^T=
 \gamma_{D2}\cdot
(C_{L}^{\mu e} ,\,   C_{L3}^{\mu e} , \,    C_{R}^{\mu e})^T
$,
where $\gamma_{D2}$ is 
\be
 \gamma_{D2} =
\frac{N_c y_t^2}{(16 \pi^2)^2} 
\left( \begin{array}{ccc}
	0 & 0 & -3/2  \\[.2cm]
	y_e/y_\mu   & -y_e/y_\mu& 1/2
\end{array}\right) \approx 
\frac{N_c y_t^2}{(16 \pi^2)^2} 
\left( \begin{array}{ccc}
	0 & 0 & -3/2  \\[.2cm]
	0 & 0 & 1/2
\end{array}\right)\, . 
\label{use2}
 \ee
We note that the  contribution of $\O_L$ and $\O_{L3}$  to $\O_{DB}$ is proportional to $2 Y_{e_R}+Y_H=0$, which seems to be an accident.

\subsubsection{Higgs quartic $\lambda$ contributions}

Starting with  the mixing from $\O_L$ to the dipoles, we find two three-particle cuts:
\be
\begin{minipage}[h][2.5cm][t]{0.31\linewidth}  \QuarticOne \end{minipage}+\quad
\begin{minipage}[h][2.7cm][t]{0.32\linewidth}  \QuarticTwo \end{minipage}
   \label{diagQ1}
\ee
The amplitude on the first cut (left diagram) is given by
\be
M_3(24_k; x_ay_bz_c) =  \sqrt{2}\lambda  Y_H \left( \frac{[x4]}{[x2][42]} {\cal C}_{kc}^{ab}   +  \frac{[yz]}{[y2][z2]} {\cal C}_{ck}^{ba}   +  \frac{[y4]}{[y2][42]} {\cal C}_{kc}^{ba}  +  \frac{[xz]}{[x2][z2]} {\cal C}^{ab}_{ck}  \right)  \, ,  \label{amp3}
\ee
where the tensors are given by ${\cal C}_{kc}^{ab}=g^\prime \cdot 2 \delta_{k}^a \delta_{c}^b$. 
 The form factor is
\be
F_{\O_L^{e\mu}}(x_ay_bz_c 1_l 3)= y_\mu  \left( \expec{13}+ \frac{2[xz]\expec{z1}}{[3x]} \right) {\cal D}_{bla}^c+ y_\mu  \left( \expec{13}+ \frac{2[yz]\expec{z1}}{[3y]} \right)  {\cal D}_{alb}^c \, ,  \label{ff3}
\ee
where ${\cal D}_{bla}^c=(- \delta^c_b \delta^{l'}_l) \eps_{l'a}$.
We are introducing these tensors in anticipation of the generalization to non-abelian SU(2) dipoles. 
It is now straightforward to compute the three-particle cut. We have to compute
$
\expec{42}[42]/((16\pi^2)^2 2! ) \int d \Omega_3  M_3(24_k; x_ay_bz_c) F_{\O_L^{e\mu}}(x_ay_bz_c 1_l 3)
$
where note that the symmetry factor $1/2!$ arises because two identical Higgses cross the cut. 
After performing the rotations in \reef{rot1} and a few elementary integrals we are led to
\be
\text{cut 1}=  \frac{12 \sqrt{2}  \lambda y_\mu g^\prime  Y_H}{(16\pi^2)^2 } \frac{\expec{12}}{[32]}  s_{24} \left[  1+ 2 \frac{s_{34}}{s_{32}}  + 2 \frac{s_{34}}{s_{32}}  \left( \frac{s_{34}+ s_{32}}{s_{32}}\right) \ln  \left(  \frac{s_{34}}{s_{34}+s_{32}} \right)  \right] \eps_{lk} \, ,  \label{log1}
\ee
where $s_{ij}=(p_i+p_j)^2$.
Regarding the second cut (right diagram of \reef{diagQ1}), we need the following amplitude
\be
M_4(34_k; x_ay_bz_c)=  \lambda y_\mu \frac{1}{[x3]} {\cal F}_{akc}^b \, ,  \label{amp4}
\ee
where ${\cal F}_{akc}^b=2 (\delta^b_k \eps_{ac} + \delta^b_c \eps_{ak})$, and the tree-level form factor
\be
F_{\O_L^{e\mu}}(x_ay_bz_c21_l) = 2 \sqrt{2} Y_{\mu_R} \frac{[xy][xz]\expec{yz}}{[12][x2]} {\cal G}_{lb}^{ac} +2 \sqrt{2} Y_{H} \frac{[xy][xz]\expec{1x}}{[y2][z2]} {\cal G}_{bl}^{ca}  \, , \label{ff4}
\ee
where ${\cal G}_{lb}^{ac}=g^\prime (- \delta_l^a \delta_b^c) $.
Once more, plugging this amplitude and form factor into \reef{rot0} (there is no $1/2!$ symmetry factor to be included for this cut), performing the rotations and a few simple integrals,  we   get
\be
\text{cut 2}=   \frac{12 \sqrt{2}  \lambda y_\mu g^\prime Y_H}{(16 \pi^2)^2} \frac{\expec{12}}{[32]} s_{34}  
 \left[  1- 2 \frac{s_{13}}{s_{32}}+  2\frac{s^2_{13}}{s_{32}^2}  \ln   \left(   \frac{s_{13}+s_{32}}{s_{13}} \right)     \right] \eps_{lk} \, .  \label{log2}
\ee
Each individual cut is  non-local, i.e. non-polynomial in the momenta. However,  the two-loop contribution to the r.h.s. of \reef{chw} must be a local contact interaction, because the one-loop mixing $\O_{L}\rightarrow \O_{DB}$ is zero. Indeed, after adding the two cuts 
\be
\text{cut 1}+\text{cut 2} = \underbrace{\phantom{\Big(}2\sqrt{2} y_\mu g^\prime \expec{12}  \expec{23} \eps_{lk} }_{ \text{dipole}} \frac{6\lambda Y_H}{(16 \pi^2)^2} \, , 
\ee
we find that the logs nicely cancel. 
Therefore the contribution of $\O_{L}^{e \mu }$ to $\O_{DB}^{e \mu }$   proportional to $\lambda$ is  given by
 $
6 \lambda Y_H/(16\pi^2)^2
$.

Next we  provide the details of the $\O_R^{e \mu}\rightarrow \O_{DB}^{e\mu }$ mixing. We need to consider two three-particle cuts:
\be
\begin{minipage}[h][2.5cm][t]{0.31\linewidth}  \QuarticThree \end{minipage}+\quad
\begin{minipage}[h][2.7cm][t]{0.32\linewidth}  \QuarticFour \end{minipage}
   \label{diagQ2}
\ee
The amplitude of the first cut is given by \reef{amp3}.  
 The form factor is
\be
F_{\O_R^{e\mu}}(x_ay_bz_c 1_l 3)= -y_e  \left( \expec{31}+ \frac{2[xz]\expec{z3}}{[1x]} \right) {\cal D}_{bla}^c - y_e  \left( \expec{31}+ \frac{2[yz]\expec{z3}}{[1y]} \right)  {\cal D}_{alb}^c \, ,  \label{ff5}
\ee
where ${\cal D}^c_{bla}=\delta^c_b \delta^{l'}_l \eps_{l'a}$.
For the second cut, the amplitude is
\be
M_5(1_l4_k; x y_bz_c)=  -\lambda y_e \frac{1}{[1x]} {\cal F}^b_{lkc}\,, \label{amp6}
\ee
where  ${\cal F}^b_{lkc}=2 (\delta^b_k \eps_{lc} + \delta^b_c \eps_{lk})$.
Note that there is a sign in \reef{amp6} w.r.t. \reef{amp4} due to crossing; this sign is crucial to get a local result, i.e. that the logs like \reef{log1} and \reef{log2} cancel. 
The form factor is given by
\be
F_{\O_R^{e\mu}}(x y_b z_c 3 2)= 2\sqrt{2}Y_{\mu_R}  \frac{[xz][xy]\expec{yz}}{[32][x2]}{\cal G}^c_b   +   2\sqrt{2} Y_H \frac{[xz][xy]\expec{3x}}{[y2][z2]}  {\cal G}^c_b\,, \label{ff6}
\ee
where ${\cal G}^c_b = g^\prime \delta^c_b$. Finally we generalize our computations for the rest of $\bar \psi \gamma \psi HDH \rightarrow \text{dipole}$ mixings. 
It  is now a simple matter to do so because the various computations differ only on the ${\cal C}\times{\cal D}$ and ${\cal F} \times {\cal G}$ tensors, which are given in Appendix~\ref{tens}. 
All in all we find that the quartic contribution to $
( \gamma_{C_{DB}^{e\mu}} , \,     \gamma_{C_{DW}^{e\mu}}  )^T=
 \gamma_{D3}\cdot
(C_{L}^{e\mu} ,\,   C_{L3}^{e\mu} , \,    C_{R}^{e\mu})^T$  is given by
\be
\gamma_{D3}=\frac{\lambda}{(16 \pi^2)^2} \left( \begin{array}{ccc}
  3 & 3 & 3 y_e/y_\mu  \\[.2cm]
  1 & 3 & y_e/y_\mu
\end{array}\right)
\approx
\frac{\lambda}{(16 \pi^2)^2} \left( \begin{array}{ccc}
  3 & 3 & 0  \\[.2cm]
  1 & 3 & 0
\end{array}\right) \,, \label{use3}
\ee
while  the quartic contribution to  $
( \gamma_{C_{DB}^{\mu e}} , \,     \gamma_{C_{DW}^{\mu e}}  )^T=
 \gamma_{D4}\cdot
(C_{L}^{\mu e} ,\,   C_{L3}^{\mu e} , \,    C_{R}^{\mu e})^T
$  is given by
\be
\gamma_{D4}=\frac{\lambda}{(16 \pi^2)^2} \left( \begin{array}{ccc}
  3 y_e/y_\mu & 3 y_e/y_\mu & 3  \\[.2cm]
  y_e/y_\mu & 3 y_e/y_\mu & 1
\end{array}\right)
\approx
\frac{\lambda}{(16 \pi^2)^2} \left( \begin{array}{ccc}
  0 & 0 & 3  \\[.2cm]
  0 & 0 & 1
\end{array}\right) \, . \label{use4}
\ee

\subsection{Finite one-loop contributions at  the electroweak scale}

Following the EFT approach  we have to  integrate out the $W$, $Z$, $h$ and top at the electroweak scale $\sim m_W$, 
and  match with the Wilson coefficients of the EFT of photons and light fermions.
In this process extra finite contributions to $d_{\mu e,e\mu}$ may be generated, as can be found in~\cite{Crivellin:2013hpa}.
We are interested in one-loop corrections arising from  $C_{L,L3,R}$ that can compete with
our two-loop calculation to the anomalous dimension of $C_{DW,DB}$. 
These are one-loop diagrams involving  a $W$ or a  $Z$ (the Higgs contributions are suppressed by extra Yukawa couplings) 
in which    $C_{L,L3,R}$ enters  via the vertex corrections  \reef{unitarygauge}.
From the $W$ we get (neglecting terms proportional to $m_e$)
\be
\Delta d_{e \mu}(m_W) = \frac{e}{32\pi^2}\frac{5}{3}C^{e \mu}_{L3}  \frac{v^2}{\Lambda^2}\,,  \label{match1}
\ee
while from the $Z$ we get
\bea
\Delta d_{e \mu}(m_W)&=&
  -\frac{e}{16\pi^2}\frac{1}{3} (C^{e \mu}_L+C^{e \mu}_{L3})\Big[ \frac{5}{4} -(\frac{1}{4}-s^2_{\theta_W})\Big] \frac{v^2}{\Lambda^2}\,,\       \nonumber \\
 \Delta d_{ \mu e}(m_W) &=&  +\frac{e}{16\pi^2}\frac{1}{3} C^{\mu e}_R\Big[ \frac{5}{4} +(\frac{1}{4}-s^2_{\theta_W})\Big]\frac{v^2}{\Lambda^2}    \,.
\label{match2}
\eea
Notice that since $s^2_{\theta_W}\approx 1/4$, these contributions  roughly cancel for BSM theories that generate
$C^{ e\mu }_L=C^{e \mu}_{L3}\not=0$, as occurs in certain models that we will discuss later.

\section{$\mu\to eee$ in the SM EFT}
 \label{mueg}

The process $\mu\to eee$ arises from the Lagrangian terms
\be
\begin{split}
\nn
-\frac{4G_F}{\sqrt{2}}\Big[&
g_1(\bar{\mu}_Re_L)(\bar{e}_Re_L)+ g_2(\bar{\mu}_Le_R)(\bar{e}_Le_R) + g_3(\bar{\mu}_R\gamma_\mu e_R)(\bar{e}_R\gamma_\mu e_R) +g_4(\bar{\mu}_L\gamma^\mu e_L)(\bar{e}_L\gamma_\mu e_L)  \\
&+ 
 g_5(\bar{\mu}_R\gamma^\mu e_R)(\bar{e}_L\gamma_\mu e_L)
+ g_6(\bar{\mu}_L\gamma^\mu e_L)(\bar{e}_R\gamma_\mu e_R)
\Big]+\text{h.c.},
\label{eq:4flowmu3e}
\end{split}
\ee
apart from the dipoles $d_{\mu e}$ and $d_{e \mu}$,  that generate the branching ratio \cite{Kuno:1999jp}
\bea
\nn
	\text{BR}\left(\mu\rightarrow eee\right)& =& 2\left(\abs{g_3}^2+\abs{g_4}^2 \right)+\abs{g_5}^2+\abs{g_6}^2+32e^2\left(\ln(\frac{m_\mu^2}{m_e^2})-\frac{11}{4}\right)(\abs{d_{\mu e}}^2+\abs{d_{e\mu}}^2)\\
	&+&8e\,\text{Re}\left(d^*_{e\mu}g_6^*+d_{\mu e}g_5^*\right)+16e\, \text{Re}\left(d^*_{e\mu}g_4^*+d_{\mu e}g_3^*\right)+\frac{1}{8}\left(\abs{g_1}^2+\abs{g_2}^2\right)\,.
	\label{BRmu333}
\eea
The Wilson coefficients entering into the $g_i$ at the tree level are
\be
\begin{split}
g_3&=-\frac{v^2}{2\Lambda^2}\Big(C_{RR}^{\mu e e e}+2 s_{\theta_W}^2C_{R}^{\mu e}\Big)\,, \ \ \
g_4=-\frac{v^2}{2\Lambda^2}\Big(C_{LL}^{\mu e e e}-(1-2s_{\theta_W}^2)\left(C_{L}^{\mu e}+C_{L3}^{\mu e}\right)\Big)\,,\\
g_5&=-\frac{v^2}{2\Lambda^2}\Big(C_{RL}^{\mu e e e}
-(1-2s_{\theta_W}^2)C_{R}^{\mu e}\Big)\,, \ \ \
g_6=-\frac{v^2}{2\Lambda^2}\Big(C_{LR}^{\mu eee}+ 2s_{\theta_W}^2\left(C_{L}^{\mu e}+C_{L3}^{\mu e}\right)\Big)\,,
\label{Wmu3e}
\end{split}
\ee
where $g_{1,2}$ are only induced by dimension-eight operators. 
Notice that in \eq{Wmu3e} all the Wilson coefficients are associated to operators with $n=4$ and $h=0$.
Furthermore, 
we see  that $C^{\mu e}_{L,L3}$ only enter in the combination $(C^{\mu e}_{L}+C^{\mu e}_{L3})$,
as  they  induce  $\mu\to 3e$  through the $Z\mu e$ coupling of \eq{unitarygauge}.

At the loop level, other Wilson coefficients can mix into  the ones in  \eq{Wmu3e}.
In particular, four-fermion $h=0$ interactions involving other families  or the combination $(C_{L}^{\mu e}-C_{L3}^{\mu e})$ 
 can enter into the RGE of the Wilson coefficients  of \eq{Wmu3e}
at the one-loop level via gauge interactions,
as they are  also $n=4,h=0$ terms and the mixing is allowed by the $\Delta n\geq |\Delta h|$ selection rule.
The corresponding RGEs are given in the Appendix \ref{app1} and the bounds obtained will be discussed later in 
 Section~\ref{constADM}.

On the other hand,  $C_{LuQe}$, $C_{LeQu}$, associated to four-fermion $|h|=2$ interactions,
and  $C^{\mu e}_{y}$, related  to an operator with $n=5$ and $|h|=1$,
cannot mix  at the one-loop level with the Wilson coefficients of  \eq{Wmu3e}, and can only 
enter into the $\mu\to 3e$ observable by mixing  into $d_{\mu e,e\mu}$
as  we explained 
for the $\mu\to e\gamma$ case.

\section{$\mu N \rightarrow e N$ in the SM EFT}
\label{muNeN}

The conversion $\mu \rightarrow e$ in nuclei  can arise from the four-fermion terms 
\be
-\frac{4G_F}{\sqrt{2}}\Big[g_{L,V}^u(\bar\mu_L\gamma^\mu e_L)(\bar u\gamma_\mu u)
+g_{R,V}^u(\bar\mu_R\gamma^\mu 	e_R)(\bar u\gamma_\mu u)
	 + g_{L,S}^u(\bar\mu_L e_R)(\bar u u)+g_{R,S}^u(\bar\mu_R e_L)(\bar u u) +(u\to d)\Big] +\text{h.c.}\,,
\label{eq:4flowmuNeN}
\ee
defined at the nuclei scale. Also the dipoles $d_{\mu e}$ and $d_{e\mu}$ can enter into this observable via the photon
splitting into quarks.
The branching ratio is given by~\cite{Kitano:2002mt}
\be
\begin{split}
	\text{BR}(\mu \rightarrow e)_N=&\frac{2G_F^2}{\omega_{capture}}\left[\left|D\, d_{\mu e} + g_{L,V}^{(p)}V^{(p)} + g_{L,V}^{(n)} V^{(n)} + g_{L,S}^{(p)} S^{(p)} + g_{L,S}^{(n)} S^{(n)}\right|^2\right.\\
	&\qquad  +\left. \left|D\, d^*_{e\mu} + g_{R,V}^{(p)} V^{(p)} + g_{R,V}^{(n)} V^{(n)} + g_{R,S}^{(p)} S^{(p)} + g_{R,S}^{(n)} S^{(n)}\right|^2\right]\,,
\end{split}
\label{BRmuNeN}
\ee
where $\omega_{capture}$ is the nuclear capture rate of the muon and $D$, $V^{(p,n)}$, $S^{(p,n)}$ are overlap integrals defined in~\cite{Kitano:2002mt}. We also define
\be
\begin{split}
	&g_{L/R,V}^{(p)}=2g_{L/R,V}^u + g_{L/R,V}^d, \qquad g_{L/R,V}^{(n)}=g_{L/R,V}^u + 2g_{L/R,V}^d\, ,\\ 
	&g_{L/R,S}^{(p)}=\sum_{q=u,d}G_S^{(q,p)}g_{L/R,S}^q,\qquad 
	g_{L/R,S}^{(n)}=\sum_{q=u,d}G_S^{(q,n)}g_{L/R,S}^q\,,
\end{split}
\ee
with $G_S^{(u,p)}\simeq G_S^{(d,n)}\simeq 5.1$, $G_S^{(d,p)}\simeq G_S^{(u,n)}\simeq 4.3$. We have neglected the contribution from the $s$ quark.
At tree level the Wilson coefficients entering into the effective  couplings \reef{eq:4flowmuNeN} are
\be
\begin{split}
&g_{L,S}^u=\frac{v^2}{2\Lambda^2}y_\mu C^{\mu e u u}_{LeQu}\,, \ \  
  g_{R,S}^u=\frac{v^2}{2\Lambda^2}y_\mu C^{e \mu  u u}_{LeQu}\,, \ \ 
 g_{L,S}^d=\frac{v^2}{\Lambda^2}y_\mu C_{LR}^{\mu dd e}\,, \ \ 
 g_{R,S}^d=\frac{v^2}{\Lambda^2}y_\mu C_{LR}^{e dd \mu}\,,\\
&g_{L,V}^u=-\frac{v^2}{4\Lambda^2}\Big[(C_{LL}^{\mu e uu}+ C_{LR}^{\mu e uu}) +2 g^u_Z\left(C_{L}^{\mu e}+C_{L3}^{\mu e}\right)\Big]\,, 
	g_{R,V}^u=-\frac{v^2}{4\Lambda^2}\Big[(C_{RL}^{\mu e uu}+ C_{RR}^{\mu e uu}) +2 g^u_Z
	C_{R}^{\mu e}\Big]\,,
\label{WCN}
\end{split}
\ee
and similarly for the down-sector with $u\to d$, where $g^u_Z=(\frac{1}{2}-\frac{4}{3}s_{\theta_W}^2)$ and 
$g^d_Z=(-\frac{1}{2}+\frac{2}{3}s_{\theta_W}^2)$. 

The  loop mixing into the Wilson coefficients of \eq{WCN}  follows the same pattern as for the 
 $\mu\to 3e$  case, as the main  difference is the replacement  $ee\to uu,dd$. 
 The only new ingredient is the presence of the $|h|=2$ operators with Wilson coefficients
  $C^{\mu e uu}_{LeQu}$  that can receive one-loop corrections from $C^{\mu e uu}_{LuQe}$ (and similarly for $u\to d$).
  These coefficients however can also receive large corrections at the QCD scale
  \cite{Dekens:2018pbu}  and will not be discussed any further here.

\section{Constraints from anomalous dimension mixings}
\label{constADM}

In Table~\ref{bounds} we present the bounds obtained when anomalous dimension  mixing is considered.
In \colorblock{blue} we show the bounds coming from one-loop mixings into the Wilson coefficients of the observables, while
in \colorblockr{red} are those in which the mixing is at the  two-loop level.
In \colorblockp{purple}
 we show the bounds for those  Wilson coefficients entering into the observables by a two-step one-loop mixing,
an  effect  of order \reef{typetwo}.
We remark that we are not including here any finite loop contributions, that could be larger but are also very model dependent
as there can be cancellations depending on the details of the model (see next section for particular cases). 

Let us start considering the bounds obtained from the two-loop mixings into the  dipole transitions $d_{e\mu,\mu e}$
that is the  novel part of this article.
The most   interesting bound  is on 
 the combination $(C^{\mu e}_L-C^{\mu e}_{L3})$ which, as we explained above,  has no serious constraint at the   tree level. 
Our two-loop calculation of the mixing  effect in \eq{use1} and \eq{use2}
shows that one can  get a  bound  from $\mu \to e\gamma$
of order
$\Lambda/\sqrt{C^{\mu e}_L-C^{\mu e}_{L3}}\gtrsim 24$ TeV (assuming  a running from that scale).
This is quite competitive as compared with  bounds coming from a one-loop mixing into 
$(C^{\mu e}_L+C^{\mu e}_{L3})$ where this latter is  strongly  constrained from $\mu\to 3e$ and $\mu N\to eN$.
The   two-loop mixing effect leads  to a bound only a factor $\sim 2$ smaller than that coming from the one-loop mixings.
This provides an interesting correlation between these 3 observables, in the sense that  if 
 one of them is measured in the near future, the other 2 should also be experimentally accessible.    
On the other hand, the  bound derived on $(C^{\mu e}_L+C^{\mu e}_{L3})$
from the two-loop mixing into  $\mu\to e\gamma$
is much weaker  (in part, because it is only generated from two loops involving  the Higgs, \eq{use3} and \eq{use4})
than that from $\mu\to 3e$ or $\mu N\to eN$  (see Table~\ref{bounds}).
This  makes it unfeasible to see this particular  BSM effect in  $\mu\to e\gamma$.

The other interesting  bound from a two-loop mixing into $\mu\to e\gamma$ is  for $C^{\mu e}_y$.
One gets  $\Lambda/\sqrt{C^{\mu e}_y}\gtrsim 4$ TeV that clearly overcomes the tree-level bound from
$h\to\mu e$. In fact, this result is already telling us  that 
 $\mu\to e\gamma$   constrains this branching ratio to be 
BR$(h\to\mu e) \lesssim 2\cdot 10^{-8}$, making it inaccessible at the LHC or even at future colliders.\footnote{
Bounds on $C_{LR}^{\mu ll e,\mu qq e}$ can also be obtained by a two-loop mixing into $\mu\to e\gamma$, 
but the contributions are proportional to  $y_{l,d}$, which leads to weak constraints.}

Let us now move to one-loop mixing effects.
Although this  analysis has been previously done  in the literature~\cite{Crivellin:2013hpa,Pruna:2014asa,Pruna:2015jhf,Crivellin:2017rmk,Calibbi:2017uvl}, 
we will provide here an understanding of the quality of the bounds  from  selection rules of 
operator mixings \cite{Cheung:2015aba,Elias-Miro:2014eia}.
In particular, the only mixing at the one-loop level into the dipoles ($n=4,|h|=2$ operators)
is $C^{\mu ett}_{LuQe}$ whose associated operator has  $n=4,|h|=2$.
This  operator can be induced from leptoquarks.
One gets a quite  strong bound, $\Lambda/\sqrt{C^{\mu ett}_{LuQe}}\gtrsim 304$ TeV.
Due to the presence of the Yukawa coupling $y_t$ in the RGE  (see \eq{dipoleluqe}) these effects are much smaller for other families.
By mixing into this Wilson coefficient $C^{\mu ett}_{LuQe}$, other four-fermion operators can enter into $\mu\to e\gamma$
by a two step one-loop mixing, and get a bound only slightly weaker (see \colorblockp{purple} bounds in Table~\ref{bounds}).
In this mixing the   $n=4,h=0$ operators need to  involve  a $y_t$ Yukawa coupling
 (as dictated by the only exception to  the selection rule $\Delta n\geq |\Delta h|$ \cite{Cheung:2015aba,Elias-Miro:2014eia}), 
and as a consequence  the mixing is only relevant for   Wilson coefficients involving the top.

Finally, other bounds on $\mu e ff$ operators, where $f$ can be any SM fermion of the 2nd and 3rd family,
can arise by mixing to $\mu eee,\mu euu,\mu edd$ at the one-loop level.
This can occur  via gauge interactions as all these operators are $n=4,h=0$.
As seen in Table~\ref{bounds} (for the 3rd family, but the same applies for the 2nd),
the bounds  range between $10-100$ TeV, depending  on the hypercharges of the states.

\section{Impact on UV Models}
\label{models}

As an application of our EFT analysis we would like to consider the impact of the discussed one- and two-loop anomalous dimensions for concrete BSM scenarios. In particular we will consider models with extra heavy fermions
and BSM that violate lepton universality.

\subsection{Heavy vector-like fermions}
\label{singlet}

Let us consider  a heavy vector-like fermion  with mass $M$ 
that can either be a singlet ($S$), a hypercharged $Y_E=-1$ state ($E$), or a SU(2)$_L$ doublet ($D$).
We assume that they couple to the SM  by mixing with the SM fermions:
\bea
\Delta {\cal L}_S&=&(y_{S}^{(1)} \bar{L}^{(1)}_L + y_{S}^{(2)} \bar{L}^{(2)}_L) S_R i \tau_2 H^*  + \text{h.c.}\,,   \nonumber\\
\Delta {\cal L}_E&=&(y_{E}^{(1)} \bar{L}^{(1)}_L + y_{E}^{(2)} \bar{L}^{(2)}_L) E_R  H  + \text{h.c.}\,,    \nonumber \\
\Delta {\cal L}_D&=&(y_{D}^{*(1)} \bar{e}^{(1)}_R + y_{D}^{*(2)} \bar{e}^{(2)}_R) D_L H^\dagger  + \text{h.c.}\,.
\eea
We would like to calculate their contributions to $\mu\to e\gamma$.
Following the EFT approach, we must first  integrate out these vector-like states at the scale $\Lambda=M$ and 
match these contributions with the Wilson coefficients of \reef{opJJ}-\reef{op4f2}. At tree level, we find
\bea
C^{e \mu}_L(M) &= -C^{e \mu}_{L3}(M) = + \frac{1}{4} y^{(1)}_S y^{*(2)}_S \ , \ {\rm for }\ S\,,     \nonumber \\
C^{e \mu}_L(M) &= \phantom{-}C^{e \mu}_{L3}(M) = - \frac{1}{4} y^{(1)}_E y^{*(2)}_E \ ,   \ {\rm for }\ E\,,    \nonumber \\
C^{\mu e}_R(M) &= -\frac{1}{2} y^{(1)}_D y^{*(2)}_D    \phantom{= - \frac{1}{4} y^{(1)}_E y^{*(2)}_E }     \ ,   \ {\rm for }\ D\,,
\label{CLi}
\eea
and $C_{L,R,L3}^{\mu e}=(C_{L,R,L3}^{e \mu})^*$, as well as
\bea
C_y^{e \mu}(M) &= &0 \, , \qquad \quad \quad \ \ C_y^{\mu e}(M)= 0 \ ,  \quad \qquad  \qquad  \qquad   \qquad   \ \ \ {\rm for}\ S\, ,   \nonumber \\
C_y^{e \mu}(M) &=& - y^{(1)}_E y^{*(2)}_E \, , \quad C_y^{\mu e}(M)= - (y_e/y_\mu) y^{(2)}_E y^{*(1)}_E \approx 0 \ ,  \quad \ \  {\rm for}\ E\,,   \nonumber\\
C_y^{\mu e}(M) &= &- y^{(1)}_D y^{*(2)}_D \, , \quad C_y^{e \mu}(M)= - (y_e/y_\mu) y^{(2)}_D y^{*(1)}_D \approx 0 \ ,  \quad \ \ {\rm for}\ D\, ,
\label{Cy}
\eea
and the Hermitian conjugates $(C_y)^*$, which are obtained by complex conjugation of \reef{Cy}. 

At the one-loop order, these heavy states also contribute to  the dipole Wilson coefficients of \eq{dip1}.
These can be extracted from the contributions to  $(g-2)$~\cite{Freitas:2014pua}. We find
\bea
C^{e \mu}_{DW}(M) -C^{e \mu}_{DB}(M)  &= \frac{1}{6} \frac{y^{(1)}_S y^{*(2)}_S}{16 \pi^2} \ , \quad \ \  {\rm for}\ S\,, \nonumber\\
C^{e \mu}_{DW}(M) -C^{e \mu}_{DB}(M)  &= \frac{1}{24} \frac{y^{(1)}_E y^{*(2)}_E}{16 \pi^2} \ ,   \quad \ \ {\rm for}\ E\,,   \nonumber \\
C^{\mu e}_{DW}(M) -C^{\mu e}_{DB}(M)  &= -\frac{1}{24} \frac{y^{(1)}_D y^{*(2)}_D}{16 \pi^2} \  ,    \quad \ \  {\rm for}\ D\, .
\label{Cdbdw}
\eea
The coefficients $C^{\mu e}_{DW,DB}$ for $S, E$, and $C^{e \mu}_{DW,DB}$ for $D$ are both $O(y_e/y_\mu) \approx 0$. 

Next,  we have to evolve these Wilson coefficients  from $M$ to the electroweak scale. 
In this RG evolution the Wilson coefficients  (\ref{CLi}) and (\ref{Cy}) mix at the two-loop level with $C^{e \mu,\mu e}_{DW,DB}$, in the manner explained in the previous sections. 
At the electroweak scale, we must now match the theory to the EFT with only photons and light fermions.
At the one-loop level, the dipoles can pick up finite terms $\Delta d_{e\mu}(m_W)$, which are given  by using 
 \reef{match1}-\reef{match2} and the tree-level matching coefficients in (\ref{CLi}). For instance, we obtain 
 $\Delta d_{e\mu}(m_W) = -\frac{y^{(1)}_S y^{*(2)}_S}{16 \pi^2} \frac{5}{12}$ for $S$.
All in all we find that the dipole coefficients are approximately given by
\bea
d_{e \mu}(m_W) &\simeq& \frac{e}{2} \frac{v^2}{M^2} \Bigg[ \Delta d_{e\mu}(m_W)+ (C^{e\mu}_{DW}(M)-C^{e\mu}_{DB}(M)) \left(1 - N_c y_t^2 \,  \frac{ \ln (M/m_W) }{16\pi^2} \right) \, \nn\\
&+& \left( (-N_c y_t^2 +2\lambda) \, C^{e\mu}_L(M) +N_c y_t^2 C^{e\mu}_{L3}(M)  - \frac{5}{8}{g'}^2 C_y^{e \mu}(M) \right) \, \frac{ \ln (M/m_W) }{(16\pi^2)^2} \Bigg] \, , \nn\\
d_{\mu e}(m_W) &\simeq& \frac{e}{2} \frac{v^2}{M^2} \Bigg[ \Delta d_{\mu e}(m_W)+ (C^{\mu e}_{DW}(M)-C^{\mu e}_{DB}(M)) \left(1 - N_c y_t^2 \,  \frac{ \ln (M/m_W) }{16\pi^2} \right) \,  \nn\\
&+& \left( (-2 N_c y_t^2 + 2 \lambda ) \, C^{\mu e}_R(M)  - \frac{5}{8}{g'}^2  C_y^{\mu e}(M) \right) \frac{ \ln (M/m_W) }{(16\pi^2)^2} \Bigg] \, .
\label{fullRG}
\eea
We should still run these coefficients from $m_W$ to $m_\mu$, but we will not include these effects here as they can be found elsewhere.

\eq{fullRG} shows that   the two-loop RG running can be sizable.
For instance,  for the singlet model $S$,
setting the Yukawa couplings of the heavy fermions to one $y_S^{(i)}= 1$, 
the current bound on $\mu \rightarrow e \gamma$ implies $M\gtrsim 43~\text{TeV}$.
In this case the RG contribution accounts for approximately the $20\%$ of the total magnitude of $d_{e\mu}$.
For the  doublet $D$ model   the RG contribution has a slightly larger impact. 
Setting again the heavy Yukawas $y_D^{(i)}= 1$, we find that  $M\gtrsim 54~\text{TeV}$, with  the RG contribution being a $25\%$ of the total magnitude of $d_{\mu e}$.
In general, for exponentially larger values of $M$ the RG contribution would dominate, but for relatively low values of $M$ the  importance of the RG contribution is model dependent.

\subsection{BSM with lepton universality violations}
BSM sectors that couple only to the muons of the 
SM have been recently proposed  to explain  some experimental discrepancies in the muon sector. 
A particular possibility is   the operator
\be
\frac{1}{M^2}  \bar L^{(2)}_L \tau^a\gamma^\mu L^{(2)}_L \bar Q^{(i)}_L\tau^a\gamma_\mu Q^{(i)}_L\,,
\label{ZL}
\ee
which could arise from integrating a heavy vector boson  that only couples to muons and to the $i$ family quarks.
In the presence of this lepton universality  breaking from some BSM, lepton number is not anymore automatically preserved since  the diagonalization of the SM Yukawa matrix $y_{e}$ leads, in the presence of \eq{ZL}, to  muon number violations. In particular, the operator ${\cal O}^{\mu ett}_{LL3}$ is induced with 
\be
\frac{C_{LL3}^{\mu ett}}{\Lambda^2}=\frac{U^{21}_{L_L}U^{\dagger\, i3}_{Q_L}U^{i3}_{Q_L}}{M^2}\,,
\label{CLL3}
\ee
where $U_{L_L,Q_L}$ is the left-handed rotation that diagonalizes $y_{e}$ and $y_{u}$.
If $y_{e,u}$ are roughly symmetric, we can estimate  $U^{21}_{L_L}\sim \sqrt{m_e/m_\mu}$ 
and $U_{Q_L}\sim V_{\rm CKM}$.
 Considering the constraint on the Wilson coefficient $C_{LL3}^{\mu ett}$ from $\mu\to e\gamma$, we obtain  the bound
	\be
	M\gtrsim 0.8\ {\rm TeV}\,, \  {\rm for}\ i=2\,, \ \ \ \
	M\gtrsim 60\ {\rm TeV}\,, \  {\rm for}\  i=3\,.
	\label{bZL}
	\ee
On the other hand, the operator \reef{ZL} also induces  a contribution to $b\to s\mu\mu$
of order
$C_{LL3}^{\mu \mu bs}/\Lambda^2={U^{\dagger\, i2}_{Q_L}U^{i3}_{Q_L}}/{M^2}$
that is bounded from \reef{bZL} to be
\be
	 \frac{C_{LL3}^{\mu \mu bs}}{\Lambda^2}\lesssim \frac{1}{(4\ {\rm TeV})^2}\,, \  {\rm for}\ i=2\,, \ \ \ \
	\frac{C_{LL3}^{\mu \mu bs}}{\Lambda^2}\lesssim \frac{1}{(290\ {\rm TeV})^2}\,, \  {\rm for}\ i=3\,. \ \ \ \
	\label{bbsmm}
	\ee
According to Ref.~\cite{Cornella:2021sby}, we need   $C_{LL3}^{\mu \mu bs}/\Lambda^2\sim 1/(56\ {\rm TeV})^2$ in order 
to  explain the  experimental discrepancy in $B\to K\mu\mu$.
From \eq{bbsmm} we see that $\mu\to e\gamma$  allows for the possibility  $i=2$ but not  $i=3$ (unless of course some of the above assumptions are relaxed).


\section{Conclusions}
\label{conclu}

In this work we  have analyzed the impact of  Lepton Flavor Violation processes with $\Delta L_\mu = \Delta L_e=1$
on the SM EFT.
The most stringent constraints  arise from  $\mu \rightarrow e \gamma$, $\mu \rightarrow eee$ and the transition rate $\mu N \rightarrow e N$, where   a rich program of measurements is planned  in the upcoming decade as summarized in Table~\ref{tabbb}. 
Given these spectacular prospects,  our  main goal here has been to understand  at which loop order  the different   dimension-six operators of the SM EFT  mix  into these LFV observables.
In particular,   we have argued that the current and future precision reach of $\mu \rightarrow e \gamma$ 
required the knowledge of  operator mixings at the  two-loop level.

We have shown that due to selection rules only one type of  operators enter at the one-loop level into 
$\mu \rightarrow e \gamma$, and  only two other types are doing it at the two-loop order.\footnote{There are also operators of type $\bar\psi^2\psi^2$ that enter at the two-loop level but their effects are suppressed by small Yukawa couplings.}
This is sketched  in Figure~\ref{figdiag}.
The two operators  mixing into $\mu \rightarrow e \gamma$  at the two-loop order are  $|H|^2 H\psi \psi$ and 
$H^\dagger D_\mu H \bar \psi \gamma^\mu \psi$ which at tree level induce LFV  $h$, $Z$ and $W$ couplings. 
While the  mixing from $|H|^2 H\psi \psi$ was already calculated in \cite{Panico:2018hal,EliasMiro:2020tdv},
we have presented  here  the calculation of the $H^\dagger D_\mu H \bar \psi \gamma^\mu \psi$  mixing.
In particular, we have calculated the two-loop anomalous dimensions of $C^{\mu e,e\mu}_{DW,DB}$ 
arising  from $C^{\mu e}_{L,L3,R}$.
Our task was greatly simplified  by using on-shell tools. 
In Section~\ref{twoloopd} we provided a lightning review of the on-shell methods that we used. 
Then, we   explained in detail the calculation of the two-loop mixings, 
showing how  this simply reduces to a product of tree-level amplitudes
integrated over some phase space.
We have also analyzed the operator mixing for the 
$\mu \rightarrow eee$ and  $\mu N \rightarrow e N$ observables, 
although in these cases we have shown that a one-loop analysis was enough. 

An interesting application of our analysis   has been 
to obtain a  bound on $(C_L^{\mu e}-C_{L3}^{\mu e})$ and 
$C^{\mu e,e\mu}_y$ from $\mu \rightarrow e \gamma$ that is competitive with bounds coming from other observables.
In particular we have shown that the bound on $C^{\mu e,e\mu}_y$ 
constrains  BR($h\to \mu e$) to be too small to be detected in future colliders.
This interplay between the different  bounds arising from the  different LFV precision measurements
was discussed in  Section \ref{constADM},
and   the actual bounds were shown in Table~\ref{bounds}  
where we  indicated the loop order of the mixing of each operator into the process of interest. 

Finally, we have shown a few illustrative examples of  how to use our  EFT analysis
to understand the different  BSM effects  inducing  $\mu\to e\gamma$.
In special, we have considered  models with heavy vector-like fermions and compared the 
 different contributions coming from EFT matching  at $\Lambda$ and $m_W$ with those from running.

Our main message here has been to show that  the next generation of LFV  
   precision measurements will require the knowledge of  renormalization effects at higher orders,
where on-shell methods have been shown to be extremely  suitable not only for performing calculations, but also for
understanding the patterns behind them.

\medskip
\section*{Acknowledgments}
We thank  Marc Riembau for early collaboration and interesting discussions. 
C.F. is supported by the fellowship FPU18/04733
from the Spanish Ministry of Science, Innovation and Universities.
A.P. has been partly supported by the Catalan ICREA Academia Program, and  grants  2014-SGR-1450, PID2020-115845GB-I00/AEI/ 10.13039/501100011033 and Severo Ochoa excellence program  SEV-2016-0588.

\newpage


\appendix

\section{One-loop anomalous dimensions relevant for LFV}
\label{app1}
In this Appendix we present the one-loop anomalous dimensions of the  Wilson coefficients 
that enter at tree level into the observables $\mu\to e\gamma$, $\mu\to eee$ and $\mu  N\to eN$. These results can be found, for example, at~\cite{Jenkins:2013wua,Alonso:2013hga}.
We are not  interested in self-renormalization but only in mixing effects coming from other Wilson coefficients not present at that level.
This gives us the leading order at which the coefficients enter into the LFV processes.

\subsection{$\mu \rightarrow e\gamma$}

The Wilson coefficient entering at tree level into $\mu \rightarrow e\gamma$ is the combination $(C_{DW}-C_{DB})$. 
This can only be renormalized at the one-loop level by operators with $|h|=2$.
In particular, the orthogonal combination $(C_{DW}+C_{DB})$  can mix into  $(C_{DW}-C_{DB})$.
Using 
\be
\frac{d}{d \ln \mu} C_{DW}=\frac{1}{16\pi^2}\left[\left(g^2\left(-\frac{11}{12}+\frac{1}{4}t_{\theta_W}^2\right)+N_cy_t^2\right)C_{DW}-\frac{1}{2}g^2t_{\theta_W}C_{DB}\right]\,,
\ee
\be
\frac{d}{d \ln \mu} C_{DB}=\frac{1}{16\pi^2}\left[-\frac{3}{2}g^2t_{\theta_W}C_{DW}+\left(g^2\left(-\frac{9}{4}+\frac{151}{12}t_{\theta_W}^2\right)+N_cy_t^2\right)C_{DB}\right]\,,
\ee
we derive this  mixing  to be 
\be
\frac{d}{d \ln \mu}(C_{DW}-C_{DB})=\frac{g^2}{16\pi^2}\left[\frac{2}{3}+\frac{1}{2}t_{\theta_W}-\frac{37}{6}t_{\theta_W}^2\right](C_{DW}+C_{DB})\,.
\ee
The other $|h|=2$ operators are \eq{op4f2}.
Among them, however,   ${\cal O}_{LeQu}$ leads to an amplitude ${\cal A}\sim \la le\ra \la qu\ra$ antisymmetric  under $l\leftrightarrow e$, so  it cannot  renormalize \eq{fdb} and \eq{fdw} that are symmetric under this exchange \cite{Baratella:2020lzz}.
Therefore only ${\cal O}_{LuQe}$ gives a nonzero contribution to the dipoles at the one-loop level 
(derived in detail in Section~\ref{oneloop}):

\be
\frac{d}{d \ln \mu}\left(\begin{array}{c}
	C_{D B} \\
	C_{D W}
\end{array}\right)=\frac{y_{u} N_c}{16 \pi^{2}}\left(\begin{array}{c}
	{5}/{12} \\
	-{1}/{4}
\end{array}\right) C_{LuQe}\,.
\label{dipoleluqe}
\ee
There are several   Wilson coefficients of \eq{op4f2} however
that can enter  into the renormalization of $C_{LuQe}$. Knowing these effects allows us to understand which
Wilson coefficients can renormalize the dipoles at the two-loop level with a double log.  
We have 
\be
\frac{d}{d \ln \mu} C_{LuQe}^{\mu e q q}=\frac{-g^{2}}{16 \pi^{2}}\left(3+5 t_{\theta_{W}}^{2}\right)\! C_{LeQu}^{\mu e qq} +
\frac{4y_u}{16 \pi^{2}}\! \left(C^{\mu e uu}_{RR} \! +\!\frac{y_e}{y_\mu}C^{\mu e qq}_{LL}\! -\!3\frac{y_e}{y_\mu}C^{\mu e qq}_{LL3}\! +\! \frac{y_e}{y_\mu}C^{\mu e uu}_{LR}\! +\!C^{\mu e qq}_{RL}\right)\,,
\label{RGEluqe1}
\ee 
\be
\frac{d}{d \ln \mu} C_{LuQe}^{e\mu qq}=\frac{-g^{2}}{16 \pi^{2}}\left(3+5 t_{\theta_{W}}^{2}\right)\! C_{LeQu}^{e\mu qq} +
\frac{4y_u}{16 \pi^{2}}\! \left(\frac{y_e}{y_\mu}C^{e\mu uu}_{RR}\! +\!C^{e\mu qq}_{LL}\! -\!3C^{e\mu qq}_{LL3} \! +\! C^{e\mu uu}_{LR}\!+\!\frac{y_e}{y_\mu}C^{e\mu qq}_{RL}\right)\,.
\label{RGEluqe2}
\ee
The first terms of both equations correspond to a $|h|=2$ operator, while the other terms correspond to $h=0$ operators where the helicity selection rule $\Delta n\geq |\Delta h|$ is violated due to the Higgs interchange that leads to a contribution 
$\propto y_uy_e$ \cite{Cheung:2015aba,Elias-Miro:2014eia}.

\subsection{$\mu \rightarrow eee$}

The Wilson coefficients $C^{\mu e}_{L,L3}$  enter in \eq{Wmu3e}  but only in the combination $C^{\mu e}_L+C^{\mu e}_{L3}$.
For this reason it is convenient to define
 \be
 C_{L\pm}=C_L\pm C_{L3}\,,
\label{CLpm}
  \ee
as we are only interested in the mixing from $C_{L-}$ into $C_{L+}$.
 From
\be
\frac{d}{d \ln \mu} C^{\mu e}_L= \frac{g^2}{16 \pi^{2}} \frac{4}{3}t_{\theta_W}^2Y_H^2 C^{\mu e}_L\,, \ \ \ \ \ \ 
\frac{d}{d \ln \mu} C^{\mu e}_{L3}= -\frac{g^2}{16 \pi^{2}} \frac{17}{3}C^{\mu e}_{L3}\,,
\label{rr2}
\ee
we obtain the mixing
\be
\frac{d}{d \ln \mu} 
C_{L+}=  \frac{g^2}{16 \pi^{2}}\Big[ \frac{17}{6}+ \frac{2}{3}t_{\theta_W}^2Y_H^2 \Big]C_{L-}\,,
\ee
where we neglected self-renormalization.

For the four-fermion $\mu eee$ Wilson coefficients entering into \eq{Wmu3e}, we have 
\bea
\frac{d}{d \ln \mu} C^{\mu eee}_{LL}&=&\frac{g^2}{16 \pi^{2}}\Bigg\{\frac{4}{3}Y_{L_L} t_{\theta_W}^2\Bigg[
N_c \left(2 Y_{Q_L} C^{\mu e qq}_{LL}  + Y_{u_R}C^{\mu e uu}_{LR}+Y_{d_R}C^{\mu e dd}_{LR}\right)+Y_H C^{\mu e}_{L}\Bigg]\nonumber\\
&&\ \ \ \ \ \ \ \ \ \ \  +\frac{2N_c C^{\mu e qq}_{LL3}}{3}+\frac{C^{\mu e}_{L3}}{3}\Bigg\}
\,,\nonumber\\
\frac{d}{d \ln \mu} C^{\mu eee}_{RR}&=& \frac{g^2}{16 \pi^{2}}\frac{4}{3}Y_{e_R}t_{\theta_W}^2 \Bigg[N_c \left(2Y_{Q_L}C^{\mu e qq  }_{RL} +Y_{u_R}C^{\mu e uu}_{RR} +Y_{d_R}C^{\mu e dd}_{RR}\right)
\Bigg]
\,,\nonumber\\
\frac{d}{d \ln \mu} C^{\mu eee}_{LR}&=&\frac{g^2}{16 \pi^{2}}\frac{4}{3}Y_{e_R}t_{\theta_W}^2 \Bigg[N_c 
\left(2Y_{Q_L} C^{ \mu e qq}_{LL} +Y_{u_R}C^{\mu e uu}_{LR} +Y_{d_R}C^{\mu e dd}_{LR}\right)+Y_HC_L^{\mu e}\Bigg]
\,,\nonumber\\
\frac{d}{d \ln \mu} C^{\mu eee}_{RL}&=&
\frac{g^2}{16 \pi^{2}}\frac{4}{3}Y_{L_L}t_{\theta_W}^2 \Bigg[N_c 
\left(2Y_{Q_L} C^{ \mu e qq}_{RL} +Y_{u_R}C^{\mu e uu}_{RR} +Y_{d_R}C^{\mu e dd}_{RR}\right)
\Bigg]\,,
\label{ADLR}
\eea
where we are interested in the case where $q,u,d$ corresponds to the 2nd and 3rd family, since for the 1st family these
Wilsons are already highly constrained at tree level by $\mu N\to eN$. 
Also we are interested in the projection  $C_{L}\to  C_{L-}/2$ and $C_{L3}\to -C_{L-}/2$.
The renormalization from $C^{\mu e \tau\tau}_{LL, LL3,RR,LR,RL}$ can be obtained from \eq{ADLR}
by the replacement $q,d\to \tau$,  $N_c\to 1$, $Y_{Q_L}\to Y_{L_L}$, $Y_{d_R}\to Y_{e_R}$, and
 putting to zero the contribution from $u$.

\subsection{$\mu N \rightarrow eN$}

The relevant anomalous dimensions of the Wilson coefficients $C^{\mu e  uu}_{LL,RR,LR,RL}$ of \eq{WCN} 
can be obtained from those in \eq{ADLR}  by the replacements $Y_{L_L}\to Y_{Q_L}$ and $Y_{e_R}\to Y_{u_R}$.

\section{Conventions and minimal form factors}
\label{conventions}

For the computations in Section~\ref{twoloopd}, we  work with 2-component Weyl spinors, and take all fermion fields to be right-handed. 
 Then the SM dimension-four Lagrangian is 
\bea
    \mathcal{L}_4 &=&  (D^\mu H^\dagger)(D_\mu H)  - \lambda (|H|^2-v^2/2)^2  \\
    & +&\sum_{\psi=Q,L,u,d,e} \psi^\dagger \overline{\sigma}^\mu D_\mu \psi - y_u \, \delta^j_i  \, H^\dagger_j \, Q^i \, u - y_d \, \epsilon_{ij} \, H^j \, Q^i \, d - y_e \, \epsilon_{ij} \, H^j \, L^i \, e\,,
\eea 
where  $i,j=1,2$ upper (lower) indices of the (anti-)fundamental representation of SU(2). All other indices like Lorentz, families etc. are contracted properly but not shown. We define $\sigma^\mu = (1, \vec{\sigma})$ and $\bar{\sigma}^\mu = (1, -\vec{\sigma})$. The covariant derivatives are  $D_\mu = \partial_\mu - i g (\tau^\alpha/2) A^\alpha_\mu - i g' Y B_\mu$, where  $\tau^\alpha$ are Pauli matrices and $Y$ is the hypercharge.

Next we report the minimal form factors in both 4-component Dirac and 2-component Weyl fermions. Remember that they are Fourier transforms of position-space form factors evaluated at zero momentum, as defined in the text. Starting with the dipole operators:
\bea
	F_{\O^{e\mu}_{DB}} &=& \langle 1^-_{L_k} 2^-_{e}  3_{H_m}  4^-_{B} |  {\overline{L}^{(1)}_{L,i}}   \Sigma^{\mu\nu} e^{(2)}_R  H^j  \delta^i_j  B_{\mu\nu}  | 0 \rangle 
	= \langle 1^-_k  2^-  3_m  4^- |  E^i  \sigma^{\mu\nu}  \mu  H^j  \epsilon_{ij}  B_{\mu\nu}  | 0 \rangle \nonumber\\
	&=&  2 \sqrt{2} \langle 14 \rangle \langle 42 \rangle  \epsilon_{km}  \,,\\[.2cm]
	F_{\O^{e\mu}_{DW}} &=& \langle 1^-_{L_k} 2^-_{e} 3_{H_m} 4^-_{W_\beta} |  {\overline{L}^{(1)}_{L,i}} \left(\tau^\alpha \right)^i_j  \Sigma^{\mu\nu}  e^{(2)}_R  H^j   W^\alpha_{\mu\nu}  | 0 \rangle 
	= \langle 1^-_k 2^- 3_m 4^-_\beta |  E^i \epsilon_{ii'} (\tau^\alpha)^{i'}_j  \sigma^{\mu\nu}  \mu  H^j  W^\alpha_{\mu\nu}  | 0 \rangle  \nonumber \\
	&=& 2 \sqrt{2} \langle 14 \rangle \langle 42 \rangle \epsilon_{kk'} (\tau^\beta)^{k'}_{m}\,, 
\eea
where  $\Sigma^{\mu\nu} = \frac{i}{2}[\gamma^\mu, \gamma^\nu]$ such that $(\overline{\psi}_L \Sigma^{\mu\nu} \psi_R)^\dagger = \overline{\psi}_R \Sigma^{\mu\nu} \psi_L$. Similarly in 2-component Weyl basis we define $\sigma^{\mu\nu} = \frac{i}{2}(\sigma^\mu \overline{\sigma}^\nu - \sigma^\nu \overline{\sigma}^\mu)$.  The relation between conjugated fermion doublets in Dirac and Weyl is $L^i = C \cdot \epsilon^{ij} \cdot L^*_{L,j}$ where $C$ is acting on Lorentz indices and $\epsilon = i \tau_2$ acting on SU(2) indices. 
Next,  the current-current operators:
\bea
     F_{\O^{e\mu}_L} &=& \langle 1^-_{L_k}  2^+_{L_l}  3_{H_m}  4^*_{H_n} |  \overline{L}^{(1)}_L \gamma^\mu L^{(2)}_L \,  H^\dagger i \overleftrightarrow{\partial_\mu} H  | 0 \rangle 
    = \langle 2^+_l 1^-_k 3_m 4^*_n |  - M^\dagger \overline{\sigma}^\mu E \, H^\dagger i \overleftrightarrow{\partial_\mu} H  | 0 \rangle \nonumber\\
&=& 2 \langle 13 \rangle [32]  (-\delta^k_l \delta^n_m ) \,, \\[.2cm]
     F_{\O^{e\mu}_{L3}} &=& \langle 1^-_{L_k}  2^+_{L_l}  3_{H_m}  4^*_{H_n} |  \overline{L}^{(1)}_L \gamma^\mu \tau^\alpha  L^{(2)}_L \, H^\dagger i \overleftrightarrow{\partial_\mu^{a}} H  | 0 \rangle 
     = \langle 2^+_l 1^-_k  3_m 4^*_n |  - M^\dagger \overline{\sigma}^\mu (-\tau^\alpha) E \, H^\dagger i \overleftrightarrow{\partial_\mu^a} H  | 0 \rangle\nonumber  \\ 
     &=& 
     2 \langle 13 \rangle [32]  (\tau^\alpha)^k_l (\tau^\alpha)^n_m  \,,\\[.2cm]
     F_{\O^{e\mu}_R} &=& \langle 1^+_{e}  2^-_{e}  3_{H_m}  4^*_{H_n} |  \overline{e}^{(1)}_R \gamma^\mu e^{(2)}_R \, H^\dagger i \overleftrightarrow{\partial_\mu} H | 0 \rangle 
     = \langle 1^+  2^-  3_m  4^*_n |  e^\dagger \overline{\sigma}^\mu \mu H^\dagger i \overleftrightarrow{\partial_\mu} H  | 0 \rangle \nonumber  \\
     &=& 2 [13] \langle 32 \rangle \delta^n_m   \,,  
\eea
where to distinguish the lepton flavor, we used letters $E$ and $e$ for electron doublet and singlet respectively, and $M$ and $\mu$ for muon.

\section{Tensors}
\label{tens}

Here we give the SU(2) tensors used in the main text to compute the anomalous dimensions:
{ \renewcommand{\arraystretch}{1.4} \renewcommand\tabcolsep{6.pt}
 \be
\begin{array}{L{1.8cm} | L{6cm}  L{7cm}     }  
 ${\cal C}_{kc}^{ab} \times {\cal D}_{bla}^{c} $ & $\O_{L} $ & $\O_{L3}$      \\  \hline
 $\O_{DB} $  & $g^\prime \cdot 2 \delta^a_k \delta^b_c \times (- \delta^c_b \delta^{l'}_l) \eps_{l'a}$ & $g^\prime \cdot 2 \delta^a_k \delta^b_c \times (\tau^\beta)^c_b (\tau^\beta)^{l'}_l \eps_{l'a} $\\ 
 $\O_{DW} $  & $g \cdot 2 (\tau^\alpha / 2)^a_k \delta^b_c \times (- \delta^c_b \delta^{l'}_l) \eps_{l'a}$ &    $g \cdot 2 (\tau^\alpha / 2)^a_k \delta^b_c \times (\tau^\beta)^c_b (\tau^\beta)^{l'}_l \eps_{l'a}$     \\  
 \end{array} \label{tensors2}
\ee
}
{\renewcommand{\arraystretch}{1.4}\renewcommand\tabcolsep{6.pt}
 \be
\begin{array}{L{1.8cm} | L{6cm}  L{7cm}  }  
 ${\cal F}^b_{akc} \times {\cal G}^{ca}_{bl} $ & $\O_{L} $ & $\O_{L3}$      \\  \hline
 $\O_{DB} $  & $2 (\delta^b_k \eps_{ac} + \delta^b_c \eps_{ak}) \times g^\prime (- \delta^c_b \delta^a_l)$ & $2 (\delta^b_k \eps_{ac} + \delta^b_c \eps_{ak}) \times g^\prime (\tau^\beta)^c_b (\tau^\beta)^a_l  $   \\ 
 $\O_{DW} $  & $2 (\delta^b_k \eps_{ac} + \delta^b_c \eps_{ak}) \times g \left( - (\tau^\alpha / 2)^c_b \delta^a_l \right)$ & $2 (\delta^b_k \eps_{ac} + \delta^b_c \eps_{ak}) \times g (\tau^\alpha / 2)^{b'}_b (\tau^\beta)^c_{b'} (\tau^\beta)^a_l $      \\  
 \end{array} \label{tensors3}
\ee}
while for the $\O_R$ operator:
{\renewcommand{\arraystretch}{1.4}\renewcommand\tabcolsep{6.pt}
 \be
\begin{array}{L{1.8cm} | L{4cm} || L{1.8cm} |  L{5cm}  }  
 ${\cal C}_{kc}^{ab} \times {\cal D}_{bla}^{c} $ & $\O_{R} $ & ${\cal F}^b_{lkc} \times {\cal G}^{c}_{b} $ & $\O_{R} $ \\  \hline
 $\O_{DB} $  & $g^\prime \cdot 2 \delta^a_k \delta^b_c \times \delta^c_b \delta^{l'}_l \eps_{l'a}$ & $\O_{DB} $ & $2 (\delta^b_k \eps_{lc} + \delta^b_c \eps_{lk}) \times g^\prime \delta^c_b $ \\ 
 $\O_{DW} $  & $g \cdot 2 (\tau^\alpha / 2)^a_k \delta^b_c \times \delta^c_b \delta^{l'}_l \eps_{l'a}$ & $\O_{DW} $ & $2 (\delta^b_k \eps_{lc} + \delta^b_c \eps_{lk}) \times g (\tau^\alpha / 2)^c_b $   \\  
 \end{array} \label{tensors4}
\ee} We emphasize that  for $\O_L$ and $\O_{L3}$ these tensors  should be used in (\ref{amp3},\ref{ff3}) and 
(\ref{amp4},\ref{ff4}), while for  $\O_R$ they should be used in (\ref{amp3},\ref{ff5}) and 
(\ref{amp6},\ref{ff6}).


\bibliography{mu2eref}
\bibliographystyle{utphys}

\end{document}